\documentclass[12pt,a4paper]{article}
\usepackage{amssymb}
\begin{document}

\title{\bf STATISTICAL DERIVATION OF BASIC EQUATIONS OF
DIFFUSIONAL KINETICS IN ALLOYS WITH APPLICATION TO THE DESCRIPTION
OF DIFFUSION OF CARBON IN AUSTENITE}
\renewcommand{\abstractname}{}
\author{V. G. Vaks, I. A. Zhuravlev}
\date{}
\maketitle

\begin{center}{\it National Research Center "Kurchatov Institute", 123182
Moscow, Russia} \vskip3mm {\it Moscow Institute of Physics and
Technology (State University), }

\hskip2mm{\it 117303 Moscow, Russia}
\end{center}

\begin{abstract}
\noindent Basic equations of diffusional kinetics in alloys are
statistically derived using the master equation approach. To
describe diffusional transformations in substitution alloys, we
derive the ``quasi-equilibrium'' kinetic equation which generalizes
its earlier versions by taking into account possible ``interaction
renormalization'' effects. For the interstitial alloys Me-X,  we
derive the explicit expression for the diffusivity $D$ of an
interstitial atom X which notably differs from those used in
previous phenomenological treatments. This microscopic expression
for $D$
is applied to describe the diffusion of carbon in austenite
basing on some simple models of carbon-carbon interaction. The
results obtained enable us to make certain conclusions about the
real form of these interactions, and about the scale of the
``transition state entropy'' for diffusion of carbon in austenite.
\end{abstract}

%PACS: 05.70.Fh; 05.10.Gg

\section{INTRODUCTION}

The problem of development of an adequate theoretical description of
diffusion in alloys attracts interest from both fundamental and
applied standpoints, see, e. g.
\hbox{\cite{Agren-82a}-\cite{KSSV-11}}. Presently, this description
is usually based on the phenomenological theory of diffusion in
multicomponent systems developed by Onsager many years ago
\cite{Onsager-31}. Phenomenological kinetic coefficients of this
theory are calculated using various simplified models with
parameters estimated empirically
\cite{Agren-82a}-\cite{Bhadeshia-81}. However, these empirical
models have usually no  consistent theoretical justification, and
their relation to interatomic interactions, as well as possibilities
of applications to other alloy systems, are typically not clear.

One of important problems in this field is the strong concentration
dependence of the diffusivity $D$ of carbon in austenite
\cite{Agren-82a}-\cite{Thibaux-07}. This dependence causes
complications in the kinetic analysis of various
diffusion-controlled reactions in steels \cite{Bhadeshia-81}, and
several empirical models have been suggested to describe this
dependence \hbox{\cite{Agren-82a}-\cite{Bhadeshia-81}.} However,
possibilities to employ these models for predictions of $D$ at
temperatures $T\lesssim 1000$ K (where many important phase
transformations occur but  $D$ can not be directly measured as
austenite is unstable here) or under partial substitution of iron by
other metals are, generally,  not clear \cite{Thibaux-07}.

The consistent statistical description of the diffusional kinetics
in alloys can be based on the master equation approach
\hbox{\cite{Martin-90}-\cite{KSSV-11}}. This approach enables us to
express the phenomenological kinetic coefficients, such as the
mobility  $M_{\alpha}$ of an ${\alpha}$-species atom, via
interatomic interactions in an alloy. These interactions can be
estimated using either some microscopic models
\cite{Blanter-99}-\cite{VKh-2} or  $ab$-$initio$ methods
{\cite{SF-07,Jiang-03,Domain-04}. As the level of accuracy and
reliability of $ab$-$initio$ calculations is steadily increasing,
this microscopic approach seems to be prospective for non-empirical
calculations of diffusivitiy.

At the same time, previous  considerations of diffusional kinetics
in alloys based on the master equation approach were usually
restricted by discussions of only simplest models and approximations
or some particular problems
\cite{Martin-90,VBD-95,Vaks-96,Nastar-00,Nastar-11}. On the other
hand, few more general discussions \cite{BV-98,KSSV-11} included
many not necessary complications and restrictions which can hinder
the understanding of the results. Therefore, the first aim of this
paper is to present a clear and general derivation of basic
equations of diffusion in alloys based on the master equation
approach, for both substitution and interstitial alloys.

To this end, in Sec. 2 we first discuss the vacancy-mediated
kinetics under diffusional transformations in substitution alloys.
This problem has been considered in Ref. \cite{KSSV-11}, and some
equations derived in Sec. 2 have been already presented there.
However, the full derivation of these equations has not been given
in \cite{KSSV-11}, while the similar derivation in \cite{BV-98}
includes a number of complications and inaccuracies. In Sec. 2  we
also discuss the methods of computer simulations based on the
microscopic equations proposed, including some generalizations of
the earlier-discussed ``equivalence theorem'' \cite{BV-98,KSSV-11}
which greatly simplifies such simulations. In Sec. 3 we consider the
interstitial alloys Me-X and derive the general statstical
expression for the diffusivity $D$ of an interstitial atom X which
has a simple and physically transparent form.  This expression
includes only  microscopic parameters which can be estimated using
either theoretical models or $ab$-$initio$ calculations. We also
generalize this microscopic expression for $D$ for the case of
multicomponent alloys (Me$_1$Me$_2$...)-X with several species atoms
in the metal sublattice.

The second aim of this work  is to apply the results of Sec. 3 to
microscopically treat the above-mentioned problem of diffusion of
carbon in austenite. This treatment described in Sec. 4 is based on
the microscopic model of carbon-carbon (C-C) interactions in
austenite suggested by Blanter \cite{Blanter-99} which supposes a
strong ``chemical'' repulsion at short  C-C distances $R_{\rm CC}$
and a purely deformational (or ``strain-induced'') interaction at
longer distances $R_{\rm CC}$. We show that some natural
generalizations of this model enable us to describe both
thermodynamic and diffusional characteristics of carbon in austenite
at the same level of accuracy as that achieved in phenomenological
models
\hbox{\cite{Agren-82a,Agren-82c,Agren-86,Bhadeshia-81,Agren-79}}. At
the same time, the microscopic approach used enables us to make a
number of conclusions about the type of carbon-carbon interactions
and about some physical features of diffusion of carbon in
austenite. The main results of this work are summarized in Sec. 5.

\section{EQUATIONS OF VACANCY-MEDIATED KINETICS FOR
DIFFUSIONAL TRANSFORMATIONS IN SUBSTITUTION ALLOYS}

\subsection{General equations for mean occupations of lattice sites}

First we present the necessary relations from Ref. \cite{KSSV-11}
with some extensions and  comments. We consider a substitutional
alloy with $(m+1)$ components \,$p'$\, that includes atoms of $m$
different species ${p}={p}_1,{p}_2,\ldots {p}_m$ and vacancies $v$:
\,$ p'=\{{p},{v}\}$\,. The distributions of atoms over the lattice
sites $i$ are described by the different occupation number sets
$\{n_i^{p'}\}$ where the operator $n_i^{p'}$ is 1 when the site $i$
is occupied by a \,$p'$-species component and 0 otherwise. For each
\,$i$\, these operators obey the identity \,$\sum_{p'}n_i^{p'}=1$,\,
so only \,$m$\, of them are independent. It is convenient to mark
the independent operators with Greek letters  \,$\rho$\,  or
\,$\sigma$\,: \,$(n_i^{p'})_{indep}= n_i^{\rho}$,\, while the rest
operator  denoted as $n_i^h$ is expressed  via $n_i^{\rho}$:
\begin{equation}
n_i^h=\Big(1-\sum_{\rho}n_i^{\rho}\Big).\label{n^h}
\end{equation}
Note that both  $n_i^h$ and $n_i^{\rho}$ are the projection
operators:
\begin{equation}
(n_i^h)^2=n_i^h,\quad n_i^hn_i^{\rho}=0,\quad
n_i^{\rho}n_i^{\sigma}=
\delta_{\rho\sigma}n_i^{\rho}.\label{n^h-n^rho}
\end{equation}
For dilute alloys, it is convenient to put ``$h$" in (\ref{n^h}) to
be the host component, e.g., $h= {\rm Fe}$ for the dilute BCC
Fe-Cu-$v$ alloys discussed in \cite{SF-07,KSSV-11} and used below
for illustrations.

In terms of all operators $n_i^{p'}$ the total configurational
Hamiltonian $H^t$ (for simplicity supposed to be pairwise) can be
written as
\begin{equation}
H^t={1\over
2}\sum_{p'q',ij}V_{ij}^{p'q'}n_i^{p'}n_j^{q'}.\label{H^t}
\end{equation}
After elimination of the operators $n_i^h$ according to (\ref{n^h}),
this Hamiltonian takes the form
\begin{equation}
H^t= E_0+\sum_{\rho i}\varphi_{\rho}n_i^{\rho}+H_{int},\quad
H_{int}=\sum_{\rho\sigma,
i>j}v_{ij}^{\rho\sigma}n_i^{\rho}n_j^{\sigma} \label{H_int}
\end{equation}
which includes only independent $n_i^{\rho}$, while constants $E_0$,
$\varphi_{\rho}$, and ``configurational interactions''
$v_{ij}^{\rho\sigma}$ are linearly expressed in terms of the
couplings $V_{ij}^{p'q'}$ in (\ref{H^t}), in particular:
\begin{equation}
v_{il}^{\rho\sigma}=(V^{\rho\sigma}-V^{\rho{h}}-V^{{h}\sigma}
+V^{hh})_{ij}.\label{v_il}
\end{equation}
The fundamental master equation for the probability $P$ of finding
an occupation number set $\{n_i^{\rho}\}=\xi$ is \cite{Vaks-04}:
\begin{equation}
dP(\xi)/ dt=\sum_{\eta} [W(\xi,\eta)P(\eta)-
W(\eta,\xi)P(\xi)]\equiv\hat SP\label{dP/dt}
\end{equation}
where \,$W(\xi,\eta)$\, is the $\eta\rightarrow\xi$ transition
probability per unit time. If we adopt   for probabilities \,$W$\,
the conventional ``transition state'' model \cite{SF-07}, we can
express the transfer matrix \,$\hat S$\, in  (\ref{dP/dt}) in terms
of the probability of an elementary inter-site exchange (``jump'')
p$i\rightleftharpoons {v}j$\, between neighboring sites \,$i$\, and
\,$j$:\,
\begin{equation}
W_{ij}^{ pv}=n_i^{p}n_j^{v}\omega_{pv}^{eff}\exp[-\beta(\hat
E_{{p}i,{v}j}^{SP}-\hat E_{{p}i,{v}j}^{in})]\label{W_ij^pv}
\end{equation}
where   \,$\beta =1/T$\, is the reciprocal temperature, \,$\hat
E_{{p}i,{v}j}^{SP}$\, is the saddle point energy, $\hat
E_{{p}i,{v}j}^{in}$ is the initial (before the jump) configurational
energy of a jumping atom \,$p$\, and a vacancy, and  the
pre-exponential factor \,$\omega_{pv}^{eff}$\, can be written as
\begin{equation}
\omega_{pv}^{eff}=\omega_{pv}\exp\,\big(\Delta
S_{{p}i,{v}j}^{SP}\big).\label{omega_pv}
\end{equation}
Here \,$\omega_{pv}$\, is the attempt frequency which, generally, is
expected to have the order of magnitude of a mean frequency of
vibrations of a jumping atom in an alloy, and  \,$\Delta
S_{{p}i,{v}j}^{SP}$\, is the entropy difference between the
saddle-point and the initial alloy states. This difference is mainly
due the difference of atomic vibrations in the  saddle-point state
(supposed to be locally equilibrium so that such thermodynamic
notion as the entropy can be applied to it) and the initial state.
At high temperatures $T$ under considerations (actually, already at
$T\gtrsim \theta/2$ where $\theta_D$ is the Debye temperature
\cite{LL}), this entropy difference can be described by the
classical expression:
\begin{equation}
\Delta S_{pv}^{SP}=3\ln\Big(\bar{\omega}_p^{in}/
\bar{\omega}_p^{sp}\Big) \label{Delta-S_sp}
\end{equation}
where $\bar{\omega}_p^{in}$ and $\bar{\omega}_p^{sp}$ are certain
average frequencies of vibrations of a jumping atom in the initial
and in the saddle point states, respectively, see, e. g.,
\cite{VKh-1}. As frequencies $\omega_p^{sp}$ in the saddle-point
configuration can notably soften with respect to $\omega_p^{in}$,
the entropy difference $\Delta S_{pv}^{SP}$ can  be expected to take
significant positive values. For example, for the Fe-Cu-$v$ alloys
with the Debye frequency $\omega_D^{\rm Fe}\simeq 6\cdot 10^{13}$
sec$^{-1}$ \cite{Kittel}, Soisson and Fu (SF) \cite{SF-07} found:
\,$\omega_{{\rm \, Fe}\,v}^{eff}\sim 80\,\omega_D^{\rm Fe}$,\,
$\omega_{{\rm Cu}\,v}^{eff}\sim 30\,\omega_D^{\rm Fe}$.\, It
implies: \,$\Delta S_{{\rm Fe}\,v}^{SP}\sim 4.5$,  \,$\Delta S_{{\rm
Cu}\,v}^{SP}\sim 3.5$, $\bar{\omega}_{\rm Fe}^{sp}\sim \omega_D^{\rm
Fe}/4$, $\bar{\omega}_{\rm Cu}^{sp}\sim \omega_D^{\rm Fe}/3$, though
so high values of \,$\Delta S^{SP}$\, and
$\omega_D/\bar{\omega}^{sp}$ can be somewhat overestimated due to
inaccuracies of estimates \cite{SF-07}.

The saddle point energy \,$\hat E_{{p}i,{v}j}^{SP}$,\, generally,
depends on the atomic configuration near the  \,$ij$\, bond. We
describe this dependence by the model of SF \cite{SF-07} supposing
the saddle-point energy to depend only on occupations of lattice
sites \,$l$\, nearest to the center of bond \,$ij$\, (denoted by
\,$l_{nn}^{ij}$):
\begin{equation}
\hat
E_{{p}i,{v}j}^{SP}=\sum_{{q},\,l=l_{nn}^{ij}}\varepsilon_{q}^{p}\,n_l^{q}=E_{h}^{p}+
\hat\Delta^{p}_{ij}.\label{E_sp}
\end{equation}
Here \,$E_{h}^{p}$\, is the saddle point energy  for the pure host
metal, while the operator\,$\hat\Delta^{p}_{ij}$\, describes changes
in this energy due to a possible presence of minority atoms near the
bond:
\begin{equation}
E_h^{p}=z_{nn}^b\varepsilon_{h}^{p};\qquad \hat\Delta^{p}_{ij}=
\sum_{\rho,\,l=l_{nn}^{ij}}\Delta^{p}_{\rho}n_l^{\rho}\label{E_h^p}
\end{equation}
where \,$ z_{nn}^b$\, is the total number of nearest lattice sites
\,$l$\, for each bond (being \,$ z_{nn}^b=6$\, for the BCC lattice),
\,$\Delta^{p}_{\rho}$\,=\,$(\varepsilon_{\rho}^{p}-\varepsilon_{h}^{p})$,
while \,$(\varepsilon_{\rho}^{p}$ and $\varepsilon_{h}^{p})$ are the
microscopic parameters of pairwise interactions calculated by SF
using $ab$-$initio$ methods \cite{SF-07}. Note that our definitions
of $\hat\Delta^{p}_{ij}$ and \,$\Delta^{p}_{\rho}$\, differ by sign
from those used in \cite{SF-07} and \cite{KSSV-11}.

The interaction parameters $V_{ij}^{p'q'}$, $\varepsilon_h^p$ and
$\Delta^p_{\rho}$ in Eqs. (\ref{H^t}) and (\ref{E_h^p}) can be
calculated by $ab$-$initio$ methods. For the Fe-Cu-$v$ alloys it is
illustrated by SF  \cite{SF-07}. Theoretical calculations of factors
\,$\omega_{pv}^{eff}$\, in Eq. (\ref{W_ij^pv}) are more difficult
due to the presence of the entropic factor \,$\exp(\Delta S^{SP})$
in Eq. (\ref{omega_pv}). However, values of \,$\omega_{pv}^{eff}$\,
can be estimated from experimental data about the self-diffusion and
the diffusion of isolated atoms in a host metal, as described by SF
\cite{SF-07}.

As $n_i^{\rho}$ in Eqs. (\ref{n^h}),  (\ref{H_int}), and
(\ref{dP/dt}) are the projection operators obeying Eqs.
(\ref{n^h-n^rho}), the most general expression for the probability
\,$P=P\{n_i^{\rho}\}$\, in (\ref{dP/dt}) can be written in the form
of the generalized Gibbs distribution \cite{BV-98,Nastar-00,Vaks-04}
\begin{eqnarray}
&&P\{n_i^{\rho}\}=\exp \Big[\beta\Big(\Omega+\sum_{\rho
i}\lambda_i^{\rho
}n_i^{\rho}-H_{int}-\hat{h}_{int}\Big)\Big],\label{P}\\
&&\hat{h}_{int}=\frac{1}{2}\sum_{\rho\sigma,ij}
h_{ij}^{\rho\sigma}n_i^{\rho}n_j^{\sigma}+\frac{1}{6}\sum_{\rho\sigma\tau,
ijk}h_{ijk}^{\rho\sigma\tau}n_i^{\rho}n_j^{\sigma}n_k^{\tau}+\ldots
\label{h_int}
\end{eqnarray}
Here parameters \,$\lambda_i^{\rho }$\, (which are both time- and
space-dependent, in general) can be called ``site chemical
potentials'' for \,$\rho$-species atoms; they are related to local
chemical potentials $\mu_i^{\rho}$ and $\mu_i^h$ of ${\rho
}$-species and host atoms as $\lambda_i^{\rho }=(\mu_i^{\rho }-
\mu_i^{h})$\, \cite{VZhKh-10}. \,$H_{int}$\, in (\ref{P}) is the
same as in (\ref{H_int}); the parameters $h_{i...j}^{\rho...\sigma}$
in (\ref{h_int}) (also depending on both time and space) describe
possible renormalizations of interactions; and $\Omega$ is
determined by normalization.

As discussed in detail in  \cite{Vaks-04}, for the usual conditions
of phase transformations corresponding to the absence of external
fluxes of particles or energy (that is, when the alloy is a
``closed'' but not an ``open'' statistical system), the effects of
renormalizations of interactions can be expected to be
insignificant, thus one can put in (\ref{P}):
\begin{equation}
\hat{h}_{int}=0.\label{h=0}
\end{equation}
There are at least two reasons to expect the validity of  Eq.
(\ref{h=0}) for the transformations in closed systems. First, this
relation holds both before and after the transformation. For
example, it is true before an initially equilibrated alloy is
quenched from a higher temperature $T_h$ to the lower temperature
$T_l$ corresponding to another equilibrium phase (or phases), and it
is also true after the new equilibrium state at $T=T_l$ is reached.
Therefore, there is no driving force drawing the correlation
parameters $h_{i...j}^{\rho...\sigma}$ in the distribution (\ref{P})
away from their Gibbs values $h_{i\ldots j}^{\rho...\sigma}=0$.
Second, the parameters $h_{i...j}^{\rho...\sigma}$ in the
distribution (\ref{P}) describe mainly the short-range order. After
a change of external conditions, such as temperature, this
short-range order is established relatively fast, for a time of the
order of one interatomic exchange time $\tau_a$, while the time for
completing microstructural evolution under phase transformation is
usually much longer: $t\gg\tau_a$  \cite{VBD-95,BV-98,KSSV-11}.
Therefore, possible fluctuative violations of relation (\ref{h=0})
at small \,$t \lesssim\tau_a $\, are not important for the whole
evolution.

When relation (\ref{h=0}) is obeyed, Eq. (\ref{P}) takes the form:
\begin{equation}
P\{n_i^{\rho}\}=\exp \Big[\beta(\Omega+\sum_{\rho i}\lambda_i^{\rho
}n_i^{\rho}-H_{int})\Big]\label{QGD}
\end{equation}
which will be called the ``quasi-equilibrium Gibbs distribution''.

Note, however, that for the essentially ``open'' systems, such as
alloys under irradiation \cite{VK-93,VB-94} or an alloy with an
external atomic flux imposed \cite{Nastar-00}, the quasi-equilibrium
relation (\ref{h=0}) can be significantly violated. Important cases
of such violations can be the phase transformations accompanied by
significant fluxes of vacancies into the transformation region, for
example, the precipitation in Fe-Cu-$v$ alloys where these fluxes
arise due to the strong trapping of vacancies by the copper-based
precipitates \cite{SF-07}. In such cases, the effective
vacancy-copper interactions can notably vary with the evolution
time, and the presence of significant renormalizations
$h_{ij}^{v{\rm Cu}}\neq 0$ can be expected. Possible methods of
calculations of these renormalizations have been discussed in
\cite{BV-98,Nastar-00,Nastar-11}.

Multiplying Eq. (\ref{dP/dt}) by operators $n_i^{\rho}$ and summing
over  all configurations $\{n_i^{\sigma}\}$, we obtain the set of
equations for mean occupations of sites (``local concentrations'')
\,$c_i^{\rho}=\langle n_i^{\rho} \rangle$:
\begin{equation}
dc_i^{\rho}/dt= \langle n_i^{\rho}\hat S\rangle \label{c_rho-dot}
\end{equation}
where
\,$\langle(...)\rangle=\sum_{\{n_j^{\sigma}\}}(...)P\{n_j^{\sigma}\}$\,
means averaging over the distribution \,$P$,\, for example:
\begin{equation}
c_i^{\rho}=\langle n_i^{\rho}\rangle =
\sum_{\{n_j^{\sigma}\}}n_i^{\rho}P\{n_j^{\sigma}\}. \label{c_rho}
\end{equation}

For what follows it is convenient to mark the minority atoms by
Greek indices $\alpha, \beta,...$ Then index $p$ in Eqs.
(\ref{W_ij^pv})-(\ref{P}) is $\alpha$ or $h$, corresponding to a
minority or a host atom, while index $\rho$ in Eqs.
(\ref{P})-(\ref{c_rho-dot}) is $\alpha$ or $v$, corresponding to a
minority atom or a vacancy. Then the system of equations
(\ref{c_rho-dot}) can be explicitly written as follows:
\begin{eqnarray}
\hskip-10mm&&\frac{dc_i^{\alpha}}{dt}=\sum_{\{n_k^{\rho}\}}\sum_{j(i)}
\omega_{\alpha {v}}^{eff}\Big[n_i^{v}n_j^{\alpha}\exp\big(\beta \hat
E_{\alpha i,{v}j}^{in}-\beta\hat E_{\alpha i,{v}j}^{SP}\big)-\{i\to
j\}\Big]\exp[\beta \big(\Omega-H_{eff}\big)]\nonumber\\
\hskip-10mm&&\frac{dc_i^{v}}{dt}=\sum_{\{n_k^{\rho}\}}\sum_p\sum_{j(i)}\omega_{pv}^{eff}
\Big[n_i^{p}n_j^{v}\exp\big(\beta \hat E_{pi,{v}j}^{in}-\beta\hat
E_{{p}i,{v}j}^{SP}\big)-\{i\to j\}\Big]\exp[\beta
\big(\Omega-H_{eff}\big)]\label{dc_v-dt}
\end{eqnarray}
where symbol \,$j(i)$\, means summation over sites \,$j$\, being
nearest neighbors of site \,$i$,\, and $p$ is $h$ or $\alpha$, while
$H_{eff}$=$H_{eff}\{n_k^{\rho}\}$ is the effective Hamiltonian for
statistical averaging of expressions in square brackets:
\begin{equation}
H_{eff}=-\sum_{\rho,k}\lambda_k^{\rho}n_k^{\rho}+
\frac{1}{2}\sum_{\rho\sigma,kl}\tilde{v}_{kl}^{\rho\sigma}n_k^{\rho}n_l^{\sigma},
\qquad \tilde{v}_{kl}^{\rho\sigma}=
({v}_{kl}^{\rho\sigma}+h_{kl}^{\rho\sigma}).\label{H_eff}
\end{equation}
Here $\tilde{v}_{kl}^{\rho\sigma}$ can be considered as the full
effective interaction, which for simplicity is supposed to be
pairwise, just as the ``true'' interaction ${v}_{kl}^{\rho\sigma}$
in  the hamiltonian (\ref{H_int}). The operator $\hat
E_{pi,{v}j}^{in}$ in (\ref{dc_v-dt}) (describing that part of the
initial configurational energy which depends on occupations of sites
$i$ and $j$) can be expressed in terms of formal variational
derivatives of the hamiltonian (\ref{H^t}) over $n_{{p}i}$ and
$n_{vi}$, $H_{{p}i}^t= \delta H^t/\delta n_{{p}i}$ and
$H_{{p}i,vj}^t=\delta^2 H^t/\delta n_{{p}i} \delta n_{vj}$:
\begin{equation}
\hat E_{{p}i,vj}^{in}= n_{{p}i}H_{{p}i}^t+n_{vj}H_{vj}^t-
n_{{p}i}n_{vj}H_{{p}i,vj}^t\label{E^in}
\end{equation}
where the third term corresponds to  the substraction of the
``double-counted'' interaction between an atom $p$ at site $i$ and a
vacancy $v$ at site $j$.

The main idea of further manipulations (analogous to those made in
\cite{BV-98,Nastar-00}) is to reduce the averages of complex
operators in  square brackets in (\ref{dc_v-dt}) to some more simple
averages having a clear physical meaning. To this end, in sums over
all configurations $\{n_k^{\rho}\}$ in (\ref{dc_v-dt}) we first
perform summation over all possible occupations of only two sites,
$i$ and $j$, belonging to the $ij$ bond under consideration. Due to
the presence  in (\ref{dc_v-dt}) of the projection operator
$n^{p}_in^{v}_j$, the summation reduces to putting
$n^q_i=\delta_{qp}$ and $n^{\rho}_j=\delta_{\rho v}$ in the $n^q_i$-
and $n^{\rho}_j$-dependent exponential factor $\exp Y$ multiplied by
this projection operator, where
\begin{equation}
Y=\beta (E_{pi,vj}^{in}-H_{eff}).\label{Y}
\end{equation}
Note that the saddle-point energy $\hat E_{{p i},{v}j}^{SP}$,
according to its definition (\ref{E_sp}), does not contain operators
of occupations of site $i$ and $j$. Therefore, the common factor
\hbox{$\omega_{pv}^{eff}\exp(\beta\Omega- \beta\hat E_{{p
i},{v}j}^{SP})$}  will be skipped for brevity in Eqs.
(\ref{n^alpha,v-term-1})-(\ref{n^h,v-term}).

For what follows it is convenient to restore formally  the summation
over all occupation number sets $\{n_k^{\rho}\}$ in (\ref{dc_v-dt}),
including all values of $n^{\rho}_i$ and $n^{\sigma}_j$. To this end
we can introduce the operator $n^h_in^h_j$ in the summand. Since
this projection operator is nonzero only when  all $n^{\rho}_i$ and
$n^{\sigma}_j$ are zero, the summation with this factor over all
possible occupations of sites $i$ and $j$ is equivalent to omitting
all $n^{\rho}_i$- and $n^{\sigma}_j$-dependent terms in the
exponential $\exp Y$. Therefore,  the result of the summation can be
written as
\begin{equation}
\sum_{n^{\rho}_i,n^{\sigma}_j}n^p_in^v_j\exp
Y=\sum_{n^{\rho}_i,n^{\sigma}_j}n^h_in^h_j\exp \left(Y_{{p i}} +Y_{v
j}+Y_{{p i}, v j}+Y\right)\label{Y_pi,vj}
\end{equation}
Here $Y_{{p i}}$, $Y_{v j}$ and $Y_{{p i}, v j}$ are the variational
derivatives of the operator $Y$ over the relevant occupation
numbers: $Y_{{p i}}=\delta Y/\delta n^p_i$, etc.  The first term of
the exponential in Eq. (\ref{Y_pi,vj}) corresponds to the
contribution to the sum (\ref{Y_pi,vj}) of the term in $Y$ linear in
$n^p_i$ but not in $n^v_j$, the second, to that of the term  in $Y$
linear in $n^v_j$ but not in $n^p_i$, and the third,  to that of the
term  in $Y$ linear in both $n^p_i$ and $n^v_j$.

Let us first consider in (\ref{Y_pi,vj}) the term with $p=\alpha$
and express all operators $n^h_l$ in the expression (\ref{E^in}) for
$E_{\alpha i,vj}^{in}$ via the independent $n^{\rho}_l$ using Eq.
(\ref{n^h}). After the above-mentioned putting in this expression
$n^{\alpha}_i=1, \ n^v_j=1$, the exponent $Y$ in (\ref{Y}) takes the
form
\begin{equation}
Y=\beta \Big[\sum_{l\rho}\Big(V_{il}^{\alpha\rho}-V_{il}^{\alpha
h}\Big)n_l^{\rho}+\sum_{l\rho}\Big(V_{jl}^{{v}\rho}-V_{jl}^{{v}h}\Big)n_l^{\rho}
+\sum_{l}\Big(V_{il}^{\alpha
h}+V_{jl}^{vh}\Big)-V_{ij}^{\alpha{v}}-H_{eff}\Big].\label{n^alpha,v-term-1}
\end{equation}
Using Eqs. (\ref{H_eff}) and  (\ref{v_il}) we can explicitly write
the relation  (\ref{Y_pi,vj}) at $p=\alpha$ as follows:
\begin{eqnarray}
&&\hskip-10mm\sum_{n^{\rho}_i,n^{\sigma}_j}n^{\alpha}_in^v_j\exp
Y=\sum_{n^{\rho}_i,n^{\sigma}_j}n_i^hn_j^h\exp\Big\{\beta
\Big[\sum_{l\rho}(u^{\rho}_{il}+u^{\rho}_{jl})n_l^{\rho}+\sum_{l}\Big(V_{il}^{\alpha
h}+V_{jl}^{{v}h}\Big)-V_{ij}^{hh}-h_{ij}^{\alpha v}\nonumber\\
&&\hskip65mm+
\lambda_i^{\alpha}+\Big(\lambda_j^{v}-\sum_{l\rho}h^{v\rho}_{jl}n_l^{\rho}\Big)
-H_{eff}\Big]\Big\} \label{n^alpha,v-term-2}
\end{eqnarray}
where the quantity
\begin{equation}
u^{\rho}_{il}= (V_{il}^{\rho h}-V_{il}^{hh})\label{u^rho_il}
\end{equation}
can be called ``the kinetic interaction'' for a $\rho$-species atom
(as it influences only mobilities but not thermodynamic properties
\cite{KSSV-11}). Note that the vacancy concentration $c_i^v=\langle
n^v_i\rangle$ in real substitution alloys is very small, hence all
$n^{v}_l$ can be neglected in the statistical averages which enter
Eqs. (\ref{dc_v-dt}). Therefore, terms with $n^{\rho}_l$ in
(\ref{n^alpha,v-term-2}) correspond actually to the minority atoms
with $\rho=\beta\neq v$. In writing Eq. (\ref{n^alpha,v-term-2})  we
also used the above-mentioned considerations that for the usual
conditions of phase transformations, the significant
renormalizations $h_{ij}^{v\alpha}$ can be expected only for the
vacancy-atom interactions, while  for the interactions between
different atoms, the analogous renormalizations
$h_{ij}^{\alpha\beta}$  are  not essential.

In the case $p=h$,\, \,Eq. (\ref{Y_pi,vj}) is simplified as the
operator $Y$ in (\ref{Y}) depends only on the independent operators
$n^{\rho}_j$ but not on $n^h_i$. Therefore, terms $Y_{hi}$ and
$Y_{hi,vj}$ in (\ref{Y_pi,vj}) are absent, and the exponent is
reduced to $(Y_{vj}+Y)$. Making the same manipulations as described
above, we obtain for this case instead of (\ref{n^alpha,v-term-2}):
\begin{eqnarray}
&&\hskip-10mm\sum_{n^{\rho}_i,n^{\sigma}_j}n^h_in^v_j\exp
Y=\sum_{n_i^{\rho},n_j^{\sigma}}n_i^{h}n_j^h\exp\Big\{\beta
\Big[\sum_{l\beta}(u^{\beta}_{il}+u^{\beta}_{jl})n_l^{\beta}+
\sum_{l}\Big(V_{il}^{hh}+V_{jl}^{{v}h}\Big)-V_{ij}^{hh}\nonumber\\
&&\hskip65mm+\Big(\lambda_j^{v}-
\sum_{l\beta}h^{v\beta}_{jl}n_l^{\beta}\Big)
-H_{eff}\Big)\Big]\Big\}.\label{n^h,v-term}
\end{eqnarray}

Substituting relations  (\ref{n^alpha,v-term-2}) and
(\ref{n^h,v-term}) in (\ref{dc_v-dt}) we can express the derivatives
\,$dc^{\rho}_i/dt$\, via some statistical averages. In writing these
averages we can take into account that the interaction
renormalizations $\hat h_{int}$ in (\ref{P}) are present only for
the vacancy-atom terms $h^{v\beta}_{ij}n_i^vn_j^{\beta}$ which
include the vacancy occupation operators $n_i^v$ and thus can be
neglected. Therefore, the quasi-equilibrium distribution (\ref{QGD})
can be used in calculations of these averages. To simplify formulas,
in treatment of  interaction renormalization terms
$h^{v\beta}_{jl}n_l^{\beta}$ in Eqs. (\ref{n^alpha,v-term-2}) and
(\ref{n^h,v-term}) we will use the mean-field approximation (MFA),
replacing each operator \,$n^{\beta}_l$\, in these averages by its
mean value \,$c^{\beta}_l$. Therefore, each term \,$h^{v\beta}_{jl}
n_l^{\beta}$\, is replaced by \,$h^{v\beta}_{jl} c_l^{\beta}$,\,
which corresponds to replacing the vacancy site chemical potential
$\lambda^v_i$ by its ``renormalized'' value $\tilde{\lambda}^v_i$:
\begin{equation}
\Big(\lambda^v_i-\sum_{l\beta}h^{v\beta}_{jl}n_l^{\beta}\Big)\to\tilde{\lambda}^v_i=
\Big(\lambda^v_i-\sum_{l\beta}h^{v\beta}_{jl}c_l^{\beta}\Big).\label{tilde-lambda}
\end{equation}
In Sec. 2.2 we show that  at \,$c^{\beta}_l$ values close to unity,
when the interaction renormalization effects can be expected to be
most significant, the approximation (\ref{tilde-lambda}) becomes
exact. Hence, one can expect it to be sufficiently accurate at all
\,$c^{\beta}_l$. We can also note that a similar MFA treatment of
interaction renormalization effects have been used by Nastar et al.
\cite{Nastar-00,Nastar-11}, and comparing their MFA and kinetic
Monte Carlo results, these authors found the MFA accuracy to be
sufficient for treatments of the renormalization effects.

Using relations   (\ref{n^alpha,v-term-2})-(\ref{tilde-lambda}) we
can  write Eqs. (\ref{dc_v-dt}) in the concise form
\begin{eqnarray}
dc_i^{\alpha}/dt&=&\sum_{j(i)}\gamma_{\alpha{v}}b_{ij}^{\alpha}
(\xi_i^{v}\eta_j^{\alpha}-\xi_j^{v}\eta_i^{\alpha})\nonumber\\
dc_i^{v}/dt&=&\sum_{j(i)}\Big[\xi_j^{v}\Big(\gamma_
{h{v}}b_{ij}^{h}+\sum_{\beta}\gamma_{\beta{v}}b_{ij}^{\beta}\eta_i^{\beta}\Big)-\{i\to
j\}\Big] \label{c_alpha-v-dot}
\end{eqnarray}
which will be called the ``quasi-equilibrium'' kinetic equations
(QKE). The term \,$\gamma_ {pv}$\, in these equations (where \,$p$\,
is $\alpha$ or $h$, i. e. denotes a minority or host atom) is the
effective exchange rate \,$p\rightleftharpoons$v\, for a pure host
metal. This term can be written in the form similar to Eq.
(\ref{W_ij^pv}):
\begin{equation}
\gamma_{pv}=\omega_{pv}^{eff} \exp\,(-\beta E_{ac}^{pv})
\label{gamma^pv}
\end{equation}
where \,$\omega_{pv}^{eff}$\,  is the same as  in (\ref{omega_pv}),
while \,$E_{ac}^{pv}$\, is the effective activation energy which is
expressed via the saddle point energies \,$E_{h}^{p}$\, in
(\ref{E_h^p}) and interactions \,$V_{ij}^{p'q'}$\, and
$h_{ij}^{\alpha v}$ in (\ref{H^t}) and  (\ref{H_eff}) as follows:\,
\begin{eqnarray}
&&E_{ac}^{\alpha{v}}=E_{h}^{p}-\sum_j(V_{ij}^{{p}h}+
V_{ij}^{{v}h})+V_{nn}^{hh}+h_{ij}^{\alpha v}\label{E_ac-alpha-v}\\
&&E_{ac}^{h{v}}=E_{h}^{h}-\sum_j(V_{ij}^{{h}h}+
V_{ij}^{{v}h})+V_{nn}^{hh} \label{E_ac-hv}
\end{eqnarray}
where $nn$ means ``nearest neighbors''. Comparing these expressions
to the analogous activation energies \,$E_{ac,{MC}}^{pv}$\, used in
the kinetic Monte Carlo approach \cite{SF-07} and given by Eq. (2.5)
in \cite{LBS-02}, we find:
\begin{equation}
E_{ac}^{\alpha{v}}=E_{ac,{MC}}^{\alpha{v}}+
\tilde{v}^{\alpha{v}}_{nn};\qquad E_{ac}^{h{v}}=E_{ac,{MC}}^{h{v}}
\label{E_ac-MC}
\end{equation}
where $\tilde{v}^{\alpha{v}}$ is the same as in (\ref{H_eff}). The
difference between $E_{ac}^{\alpha{v}}$ and
$E_{ac,{MC}}^{\alpha{v}}$ arises because in the statistically
averaged QKE (\ref{c_alpha-v-dot}), the transition probability
(\ref{W_ij^pv}) is averaged over the distribution (\ref{P}). For the
inter-site exchange $\alpha i\rightleftharpoons {v}j$,\, it leads to
an extra Gibbs factor \,$\exp (-\beta \tilde{v}^{\alpha{v}}_{ij})$\,
with \,$\tilde{v}^{\alpha{v}}_{ij}$\, from (\ref{H_eff}) in the
averaged probability.

In this connection we note that in their study of diffusion in
dilute Fe-Cu alloys \cite{SF-07} with a thorough taking into account
the correlation effects for this vacancy-mediated diffusion
\cite{Le_Clair-70}, SF obtained for an effective activation energy
of a copper atom  the value $(E_{ac}^{{\rm Cu}{v}})_{eff}\simeq$0.47
eV \cite{LBS-02,Soisson-08}. This is very close to the value
$E_{ac}^{{\rm Cu}{v}}$=0.44 eV which follows from Eq.
(\ref{E_ac-alpha-v}) at $h_{ij}^{{\rm Cu} v}=0$ after substitution
of couplings $V_{ij}^{pq}$ used by SF. It may imply that  for the
diffusion in dilute Fe-Cu alloys, our statistical averaging with the
quasi-equilibrium Gibbs distribution (\ref{QGD}) can rather
accurately describe the correlation effects mentioned
\cite{Le_Clair-70}, while the interaction renormalization effects
for this diffusion are not very significant.

The quantities  \,$b_{ij}^{p}$\, in (\ref{c_alpha-v-dot}) (to be
called ``correlators'') are certain averages of site occupations
which describe influence of minority atoms in vicinity of the bond
\,$ij$\, on the $pi$$\rightleftharpoons$$vj$  jump probability:
\begin{equation}
b_{ij}^{p}=\langle n_i^hn_j^h\exp\Big[\sum_{\alpha
l}\beta(u_{il}^{\alpha}+u_{jl}^{\alpha})n_l^{\alpha}-
\sum_{\alpha,\,l=l_{nn}^{ij}}\beta\Delta^{p}_{\alpha}n_l^{\alpha}\Big]\rangle\label{b_ij^p}
\end{equation}
where \,$\Delta^{p}_{\alpha}$\, is the same as in (\ref{E_h^p}), and
\,$u_{il}^{\alpha}$\, is the same as in
(\ref{n^alpha,v-term-2})-(\ref{n^h,v-term}).

Finally, the quantities    \,$\xi_i^{v}$\, and \,$\eta_i^{\alpha}$\,
in   (\ref{c_alpha-v-dot}) can be called the ``site thermodynamic
activities'' for   vacancies and \,$\alpha$\,-species atoms,
respectively, because they are related to the site chemical
potentials \,$\lambda_i^{\alpha}$\, in (\ref{QGD}) and the
renormalized  site chemical potential $\tilde{\lambda}_i^{v}$ in
(\ref{tilde-lambda}) as
\begin{equation}
 \xi_i^{v}=\exp\,(\beta \tilde{\lambda}_i^{v}); \qquad
\eta_i^{\alpha}=\exp\,(\beta \lambda_i^{\alpha}),\label{xi_i-eta_i}
\end{equation}
that is, similarly to the relations between conventional
thermodynamic activities  and chemical potentials.

\subsection{Calculations of site chemical potentials  \,$\lambda_i^{\rho}$\,
and correlators \,$b_{ij}^{p}$}

To find explicit expressions for site chemical potentials
\,$\lambda_i^{\rho}=\lambda_i^{\rho}(c_j)$\, determined by Eqs.
(\ref{c_rho}), and for correlators \,$b_{ij}^{p}=b_{ij}^{p}(c_k)$\,
determined by Eqs. (\ref{b_ij^p}), we should use some approximate
method of statistical physics, such as the MFA or cluster methods
\cite{Vaks-04}. As discussed in detail in
\cite{KSSV-11,VZhKh-10,VKh-1}, employing the MFA for calculations of
chemical potentials \,$\lambda_i^{\rho}$\, in  real alloys often
leads to great errors, while the pair cluster approximation (PCA)
usually combines simplicity of calculations with a high accuracy,
particularly for dilute alloys. As an illustration (used also below
for interstitial alloys), we present the PCA expressions for
\,$\lambda_i^{\rho}$\, in a binary A-B-$v$ alloy with host atoms
$h$=A and minority atoms $\alpha$=B\, (skipping index B at
\,$\lambda_i^{\rm B}=\lambda_i$\, and $c_i^{\rm B}=c_i$\,  for
brevity):

\begin{eqnarray}
&&\lambda_i=T\Big[\ln (c_i/c_i^h)+
\sum_{j\neq i}\ln (1-g_{ij}c_j)\Big] \label{lambda_PCA}\\
&&\lambda_i^{v}=T\Big[\ln (c_i^{v}/c_i^h)- \sum_{j\neq i} \ln
(1+g_{ij}^{v}c_j^{\alpha})\Big]. \label{lambda_v}
\end{eqnarray}
Here the function \,$g_{ij}$\, or \,$g_{ij}^{v}$\, is expressed via
the Mayer function \,$f_{ij}=[\,\exp\, (-\beta v_{ij})-1$]\, or
\,$f_{ij}^{v}=[\,\exp\, (-\beta v_{ij}^{v{\rm B}})-1]$\, for the
 potential \,$v_{ij}\equiv v_{ij}^{\rm BB}$\, or
\,$v_{ij}^{v{\rm B}}$\, defined in (\ref{v_il}) as follows:
\begin{eqnarray}
\hskip-10mm &&g_{ij}=2f_{ij}/[R_{ij}+1+f_{ij}(c_i+c_j)]\nonumber\\
\hskip-10mm &&g_{ij}^{v}=2f_{ij}^{v}/[R_{ij}+1+f_{ij}(c_i-c_j)] \nonumber\\
\hskip-10mm
&&R_{ij}=\left\{[1+(c_i+c_j)f_{ij}]^2-4c_ic_jf_{ij}(f_{ij}+1)\right\}^{1/2}.
\label{g_ij}
\end{eqnarray}
For a multicomponent alloy  A-B$_1$-...B$_m$-$v$, the PCA methods of
calculations of site chemical potentials \,$\lambda_i^{\rho}$\, are
described in \cite{VZhKh-10}.

In calculations of correlators \,$b_{ij}^{p}$\, (\ref{b_ij^p}), we
first consider for simplicity the case of configuration-independent
saddle-point energies when differences
\,$\hat{\Delta}^{p}_{\alpha}$\, and \,$\Delta^{p}_{\alpha}$\, in
Eqs. (\ref{E_sp}), (\ref{E_h^p}) and (\ref{b_ij^p}) are zero and
correlators \,$b_{ij}^{p}=b_{ij}$\, are independent of the kind of a
jumping atom  \,$p$.\, Using Eqs. (\ref{n^h-n^rho}) and the identity
\begin{equation}
\exp\,(xn_l^{\alpha})=1+n_l^{\alpha}f(x),\qquad f(x)=(e^x-1),
\label{f_x}
\end{equation}
which follows from  (\ref{n^h-n^rho}), we can rewrite Eq.
(\ref{b_ij^p}) as
\begin{eqnarray}
\hskip-10mm&b_{ij}&=\Big\langle n_i^hn_j^h\prod_{l=1}^{k_t}
(1+\sum_{\alpha}f_l^{\alpha}n_l^{\alpha})\Big\rangle= \nonumber\\
\hskip-15mm&&\sum_{k=0}^{k_t}\hskip2mm\sum_{l_1\neq\ldots
l_k}\sum_{\alpha_1\ldots\alpha_k}\Big\langle n_i^hn_j^h
n_{l_1}^{\alpha_1}\ldots n_{l_k}^{\alpha_k}\Big\rangle
f_{l_1}^{\alpha_1}\ldots f_{l_k}^{\alpha_k} \label{b_ij-series}
\end{eqnarray}
where we set
\begin{equation}
f_l^{\alpha}=f(\beta u_{il}^{\alpha}+\beta
u_{j\,l}^{\alpha})\label{f_l^alpha}
\end{equation}
with \,$f(x)$\, from (\ref{f_x}),\,  while \,$k_t$\, in
(\ref{b_ij-series}) is the total number of sites with nonzero values
of potentials \,$(u_{il}^{\alpha}+ u_{j\,l}^{\alpha})$.\, For
example, for the nearest-neighbor or next-to-nearest-neighbor
interaction models  in a BCC lattice \cite{KSSV-11}, we have
\,$k_t=14$\, or \,$k_t=20$.\,

In finding averages in (\ref{b_ij-series}) we should consider that
the functions \,$f_l^{\alpha}$\, in Eqs. (\ref{b_ij-series}) and
(\ref{f_l^alpha})  for real alloys are typically rather large. For
example,  for the BCC \hbox{Fe-Cu-$v$} alloys considered in
\cite{KSSV-11} we have: \,$f(\beta u_1)\sim 5$\, and \,$f(\beta
u_2)\sim 1$\, (where interaction $u_1$ and $u_2$ corresponds to the
nearest and next-to nearest neighbors). Thus the main contributions
to sum (\ref{b_ij-series}) come from averages of products of many
different operators \,$n_l^{\alpha}$\, corresponding to
well-separated and weakly correlated sites \,$l$.\, In particular,
for the BCC lattice, these products (even for the nearest-neighbor
interaction model) include terms with the neighbors from first to
tenth, most often third and fourth. Correlations of occupations of
so distant sites should typically be weak. Therefore, using the
simple MFA that neglects such correlations,  in calculations of
averages (\ref{b_ij-series}) should, generally, be adequate, unlike
calculations of chemical potentials $\lambda_i$ mentioned above.

In the MFA, each operator $n^p_l$ in Eq. (\ref{b_ij-series}) is
replaced by its average value  $c^p_l$. Hence the correlator
$b_{ij}$  can be explicitly written as
\begin{equation}
b_{ij}=c_i^hc_j^h\prod_{l=1}^{k_t}(1+\sum_{\alpha}
c_l^{\alpha}f_l^{\alpha})=c_i^hc_j^h\exp\Big[\sum_{l=1}^{k_t}
\ln\Big(1+\sum_{\alpha}f_l^{\alpha}c_l^{\alpha}\Big)\Big].\label{b_ij-c_l}
\end{equation}

When the differences \,$\Delta^{p}_{\alpha}$\, in Eqs.
 (\ref{E_h^p}) and  (\ref{b_ij^p}) are nonzero, the correlator
\,$b_{ij}^{p}$\, in Eq. (\ref{b_ij^p}) can be calculated by the same
way as \,$b_{ij}$\, in (\ref{b_ij-series})-(\ref{b_ij-c_l}). The
difference arises only for sites \,$l=l_{nn}^{ij}$\, adjacent to the
\,$ij$\, bond, for which the factor \,$f_l^{\alpha}$\, defined by
Eq. (\ref{f_l^alpha}) is replaced by an analogous factor
\,$f_{l\Delta}^{\alpha p}$\, defined as
\begin{equation}
f_{l\Delta}^{\alpha p}=f(\beta u_{il}^{\alpha}+\beta
u_{j\,l}^{\alpha}-\beta\Delta^{ p}_{\alpha l}
\delta_{l,l_{nn}^{ij}})\label{f_l-Delta^alpha-p}
\end{equation}
where $\Delta^{ p}_{\alpha l}=\Delta^{
p}_{\alpha}\delta_{l,l_{nn}^{ij}}$ and  $\delta_{l,l_{nn}^{ij}}$ is
unity when \,$l=l_{nn}^{ij}$ and zero when \,$l\neq l_{nn}^{ij}$.\,
Therefore, the correlator \,$b_{ij}^{p}$\, is given by Eq.
(\ref{b_ij-c_l}) with replacing each $f_l^{\alpha}$ by
$f_{l\Delta}^{\alpha p}$:
\begin{equation}
b_{ij}^p=c_i^hc_j^h\exp\Big[\sum_{l=1}^{k_t}
\ln\Big(1+\sum_{\alpha}f_{l\Delta}^{\alpha
p}c_l^{\alpha}\Big)\Big].\label{b_ij-Delta_p-c_l}
\end{equation}

Finally, let us make a remark about the MFA-type approximation
(\ref{tilde-lambda}) used above in the derivation of QKE
(\ref{c_alpha-v-dot}). If we don't use this approximation, Eqs.
(\ref{c_alpha-v-dot}), instead of the correlators \,$b_{ij}^{p}$\,
(\ref{b_ij-Delta_p-c_l}),  include similar correlators
\,$\tilde{b}_{ij}^{p}$\, differing from \,$b_{ij}^{p}$\,  by the
presence of additional interaction renormalization terms \,$\beta
h^{v\beta}_{jl} n_l^{\alpha}$\, in the exponents:
\begin{equation}
\tilde{b}_{ij}^p=c_i^hc_j^h\exp\sum_{l=1}^{k_t}
\ln\Big\{1+\sum_{\alpha}\Big[\exp\Big(\beta u_{il}^{\alpha}+\beta
u_{j\,l}^{\alpha}-\beta\Delta^{p}_{\alpha l} -\beta
h^{v\alpha}_{jl}\Big)-1\Big]c_l^{\alpha}\Big\}\label{tilde_b_ij^p}
\end{equation}
where we write the function $f(x)$  from (\ref{f_x}) explicitly. For
the diffusional transformations under consideration, the interaction
renormalization effects seem to be most significant in those space
regions where the local concentration $c_l^{\alpha}$ is close to
unity. It is illustrated by the case of precipitation in Fe-Cu
alloys where these effects arise due to the strong trapping of
vacancies by the Cu-based precipitates for which  $c_l^{\rm Cu}$ is
close to unity \cite{KSSV-11}. For such $c_l^{\alpha}$, the argument
of logarithm in  (\ref{tilde_b_ij^p}) is reduced to the single
exponent, and the relation (\ref{tilde-lambda}) becomes exact. Let
us also note that for the simulations of diffusional transformations
based on Eq. (\ref{c_alpha-v-dot}), the details of vacancy
distributions are actually insignificant due to the ``adiabaticity
principle'' and the ``rescaling of time'' procedure discussed in
Sec. 2.3. Therefore, the approximation (\ref{tilde-lambda}) appears
to be sufficient for using in such simulations.

\subsection{Reducing kinetic equations
(\ref{c_alpha-v-dot}) to those for some direct exchange model}

The QKE  (\ref{c_alpha-v-dot}) can be used for modeling of most
different phase transformations, in particular, of processes of
precipitation which attract great attention in connection with
numerous applications \cite{Miller-03,Rana-07,KWZS-08}. However, in
the original form  (\ref{c_alpha-v-dot}) these equations are not
suitable for using in computer simulations due to very small values
of vacancy concentration $c_v$  in real alloys. As atomic exchanges
$pi\leftrightharpoons vj$ take place only with a vacancy, this
smallness leads to a great difference in the relaxation time $\tau$
between atoms $\alpha$ and vacancies $v$: $\tau_{\alpha}\sim
\tau_{v}/c_v\gg\tau_{v}$. It is illustrated by the presence of
vacancy activities \,$\xi_i^{v}=\exp (\beta \tilde{\lambda}_i^{v
})$\, in the right-hand side of the QKE  (\ref{c_alpha-v-dot}). This
activity is proportional to the vacancy concentration \,$c_i^{v}$\,
which is  a general relation of thermodynamics of dilute solutions
illustrated by Eq. (\ref{lambda_v}). Therefore, the time derivatives
of mean occupations are proportional to the local vacancy
concentration, \,$c_i^{v}$\, or \,$c_j^{v}$.\, It is natural for the
vacancy-mediated kinetics and leads to the strong inequality between
$\tau_{\alpha}$ and $\tau_{v}$ mentioned above. Therefore, the type
of temporal evolution for atoms and vacancies is quite different,
which makes the direct numerical solving of Eqs.
(\ref{c_alpha-v-dot}) for \,$c_i^{v}(t)$\, and \,$c^{\alpha}_i(t)$\,
to be unsuitable and time consuming.

At the same time, the inequality  $\tau_{\alpha}\gg\tau_{v}$ enables
us to use the ``adiabatic'' approach used in many fields of physics,
including the well-known Born-Oppenheimer approach in the
quantum-mechanical description of motion of atoms in molecules and
solids. In this approach, the effective driving force for a slow
motion is obtained by its averaging over a rapid motion. Therefore,
to fully describe the slow motion, only few averaged characteristics
of the rapid motion are needed. In the quantum mechanics, this is
the appropriate electronic energy (``electronic term'') calculated
at the fixed positions  ${\bf R}_{i}$ of atoms. In our problem it
means that at the given atomic distribution
\,$\{c_{i}^{\alpha}\}$,\, the local vacancy concentration
\,$c_i^{v}$\, adiabatically fast (i. e., for a time $\tau_v\sim
c_v\tau_{\alpha}\ll\tau_{\alpha}$) reaches its ``quasi-equilibrium''
value \,$c_i^{v}\{ c_{ i}^{\alpha}\}$\, for which the right-hand
side of the second equation (\ref{c_alpha-v-dot}) vanishes.
Therefore, discarding small corrections of the relative order
\,$c_i^{v}\ll 1$,\, we can approximate  this equation by its
adiabatic version:
\begin{equation}
0=\sum_{j(i)}\Big[\xi_j^{v}\Big(\gamma_{h{v}}b_{ij}^{h}
+\sum_{\alpha}\gamma_{\alpha{v}}b_{ij}^{\alpha}\eta_i^{\alpha}\Big)-\{i\to
j\}\Big] \label{adiabat}
\end{equation}
which can be called ``the adiabaticity equation'' for the vacancy
activity \,$\xi_i^{v}$.\,  Solving this linear equation for
\,$\xi_i^{v}$\, we can, in principle, express it via
 \,$c_j^{\alpha}$.\, Then substitution of
these  \,$\xi_i^{v}(c_j^{\alpha})$\, into the first Eq.
\,(\ref{c_alpha-v-dot})  yields the QKE for some equivalent
direct-atomic-exchange (DAE) model.

To illustrate these considerations, we first consider the models
with the configuration-independent saddle-point energies. For such
models, the parameters \,$\Delta^{p}_{\rho}$\, in (\ref{E_h^p}) are
zero, the correlators \,$b_{ij}^{p}=b_{ij}$\,  do not depend on the
kind \,$p$\, of a jumping atom, and the adiabaticity equation
(\ref{adiabat}) takes the simple form:
\begin{equation}
\sum_{j(i)}b_{ij}\,\xi_i^{v}\xi_j^{v}
\Big[\Big(\gamma_{h{v}}+\sum_{\alpha}\gamma_{\alpha{v}}\eta_i^{\alpha}\Big)/\xi_i^{v}-\{i\to
j\}\Big]=0\label{BV-adiabaticity}
\end{equation}
If we let  $1/\nu_i$ denote the first term in the square brackets
(\ref{BV-adiabaticity}),  then the difference in these brackets
takes the form \,$(\nu_i^{-1}-\nu_j^{-1})$.\, Hence, a solution of
Eqs. (\ref{BV-adiabaticity}) is given by $\nu_i$ being a constant
independent of the site number \,$i$\, (although possibly depending
on time as well  as on temperature and other external parameters):
\begin{equation}
\nu_i=\xi_i^{v}\Big/\Big(\gamma_{h{v}}+\sum_{\alpha}\gamma_{\alpha{v}}\eta_i^{\alpha}\Big)=\nu
(t).\label{nu_i}
\end{equation}
 Relation (\ref{nu_i}) determines the above ``quasi-equilibrium'' vacancy distribution
\,$c_i^{v}\{c_i^{\alpha}\}$\, which adiabatically fast follows the
atomic distribution \,$\{c_i^{\alpha}\}$.\, Substituting it in the
first Eq. (\ref{c_alpha-v-dot}) we obtain an explicit kinetic
equation for atomic distributions \,$\{c_i^{\alpha}\}$\, for which
the evolution of the vacancy distribution is characterized by a
single parameter \,$\nu (t)$ being a ``spatially self-averaged''
quantity:
\begin{eqnarray}
&dc_i^{\alpha}/dt=&\sum_{j(i)}b_{ij}\nu(t)\Big[\gamma_{\alpha{v}}
\gamma_{h{v}}\Big(\eta_j^{\alpha}-\eta_i^{\alpha}\Big)\nonumber\\
&&+\sum_{\beta}\gamma_{\alpha{v}}\gamma_{\beta{v}}
\Big(\eta_j^{\alpha}\eta_i^{\beta}-\eta_i^{\alpha}\eta_j^{\beta}\Big)\Big].
\label{c_alpha-DAE}
\end{eqnarray}
Equations (\ref{c_alpha-DAE}) can also be rewritten in the form used
for DAE models \cite{Vaks-04}:
\begin{eqnarray}
\hskip-8mm&&dc_i^{\alpha}/dt=\sum_{j(i)}M_{ij}^{\alpha h}\,2\sinh
[\beta(\lambda_j^{\alpha}-\lambda_i^{\alpha})/2]\nonumber\\
\hskip-12mm&&+\sum_{j(i),\,\beta}M_{ij}^{\alpha\beta}2\sinh
[\beta(\lambda_j^{\alpha}+\lambda_i^{\beta}-\lambda_i^{\alpha}-\lambda_j^{\beta})/2]
\label{c_alpha-beta-sinh}
\end{eqnarray}
where the generalized mobilities \,$M_{ij}^{pq}$,\, which describe
the inter-site exchanges \,$\alpha$$\leftrightharpoons$$h$\, and
\,$\alpha$$\leftrightharpoons$$\beta$,\, are given by
\begin{eqnarray}
\hskip-12mm&&M_{ij}^{\alpha h}=\gamma_{\alpha{v}}\gamma_{h{v}}\nu(t)\,b_{ij}\exp\,[\beta(\lambda_i^{\alpha}+\lambda_j^{\alpha})/2]\label{M_ij^alpha-h}\\
\hskip-12mm&&M_{ij}^{\alpha\beta}=\gamma_{\alpha{v}}\gamma_{\beta{v}}\nu(t)\,b_{ij}\exp\,[\beta(\lambda_i^{\alpha}+\lambda_j^{\alpha}+
\lambda_i^{\beta}+\lambda_j^{\beta})/2]. \label{M_ij^alpha-beta}
\end{eqnarray}
Comparing these expressions with the expression (32) in \cite{BV-98}
which describes the mobility \,$M_{ij}^{pq}$\, in an alloy with the
nearest-neighbor direct-exchange rate
\,$\gamma_{ij}^{pq}$=$\gamma_{pq}$,\, we see that Eqs.
(\ref{M_ij^alpha-h}) and (\ref{M_ij^alpha-beta}) correspond to a DAE
model with the effective direct exchange rates
\begin{equation}
\gamma_{\alpha h}^{eff}=
\gamma_{\alpha{v}}\gamma_{h{v}}\,\nu(t);\qquad
\gamma_{\alpha\beta}^{eff}=
\gamma_{\alpha{v}}\gamma_{\beta{v}}\,\nu(t).\label{gamma_pq^eff}
\end{equation}
As $\nu(t)$ in (\ref{nu_i}) is proportional to $c^v_i$, the
effective DAE rates (\ref{gamma_pq^eff}) are by a factor \,$c_{v}$\,
smaller than the vacancy exchange rates \,$\gamma_{pv}$, in
accordance with the above-discussed adiabaticity relations.

For more realistic models with configuration-dependent saddle-point
energies, the basic adiabaticity equation (\ref{adiabat}) for
vacancy activities \,$\xi_i^{\rm v}$\, can not be solved
analytically in general, and hence either numerical or some
approximate analytic methods should be used. For the first-principle
model of Fe-Cu-$v$ alloys developed in \cite{SF-07}, such
approximate treatments made in  \cite{KSSV-11} have shown that the
equivalence relations (\ref{c_alpha-DAE})-(\ref{gamma_pq^eff})
usually preserve their form, but the correlator $b_{ij}$ is replaced
by some other quantity, $b_{ij}^{\rm Cu}$  or $b_{ij}^{\rm Fe}$.
Physically, the possibility to reduce the vacancy-mediated kinetics
to the equivalent direct atomic exchange kinetics is connected with
the above-mentioned fact that in the course of evolution of an
alloy, the distribution of vacancies adiabatically fast follows that
of the main components. Therefore, it can be assumed that this
equivalence is actually a general feature of the vacancy-mediated
kinetics, while for more general models, the correlators
\,$b_{ij}$\,  in (\ref{M_ij^alpha-h}) can be replaced by some other
expressions with similar properties.

The function \,$\nu (t)$\, in Eq. (\ref{gamma_pq^eff}) determines
the rescaling of time between the initial vacancy-mediated exchange
model and the equivalent DAE model
(\ref{c_alpha-DAE})-(\ref{gamma_pq^eff}). Temporal evolution of this
DAE model is actually described by the dimensionless ``reduced
time'' \,$t_r$\, related to the real time \,$t$\, by the
differential or integral relations
\begin{eqnarray}
&&dt_r=\gamma_{\alpha h}^{eff}dt= \gamma_{\alpha{\rm
v}}\gamma_{h{\rm v}}\nu (t)dt, \qquad t_r=\int_0^t
\gamma_{\alpha h}^{eff}(t')\,dt',\nonumber\\
&&t=\int_0^{t_r}dt_r'\tau_{\alpha h}^{eff}(t_r') \label{t-t_r}
\end{eqnarray}
where \,$\tau_{\alpha h}^{eff}=1/\gamma_{\alpha h}^{eff}$\, has the
meaning of the mean time of an atomic exchange
\,$\alpha\leftrightharpoons h$,\, while \,$t_r$\, has the meaning of
an effective number of such atomic exchanges.

The form of the function \,$t(t_r)$\, in (\ref{t-t_r}) depends on
the boundary conditions for vacancies adopted in simulations. In
particular, if we adopt the ``vacancy conservation'' model for which
the interaction renormalization effects can be expected to be
insignificant, we can use  Eqs. (\ref{nu_i}) and (\ref{lambda_v}) to
express the local vacancy concentration \,$c^v_i$\, via \,$\nu(t)$\,
and \,$c^{\alpha}_i(t_r)$.\, Then the vacancy conservation
condition: \,$\sum_ic^v_i(t,t_r)=N_v=const$,\, can be used to
explicitly find the dependence \,$t(t_r)$.\, However, taking into
account a possible creation of vacancies at various lattice defects
(grain boundaries, dislocations, etc) suggested in the kinetic Monte
Carlo (KMC) simulations \cite{SF-07,LBS-02} appears to be more
realistic. Then the dependences $t(t_r)$  can be found from
comparison of results of the DAE-based simulations described above
to the appropriate KMC results, as illustrated in \cite{KSSV-11},
and these dependences seem to be rather simple and universal. In
more detail, applications of Eqs. (\ref{c_alpha-DAE})-(\ref{t-t_r})
to studies of precipitation in concrete alloys will be described
elsewhere.
\section{EQUATIONS FOR  DIFFUSION OF
INTERSTITIAL ATOMS IN INTERSTITIAL ALLOYS}
In binary interstitial alloys Me-X where X is an interstitial atom,
in particular, in iron-carbon steels, diffusion of atoms X is
realized via thermo-activated jumps of these atoms between their
interstitial sites (``pores''). Therefore, this diffusion can be
described by the general equations of Sec. 2 for a particular case
of a substitution binary alloy X-$v$ which consists of atoms
$\alpha$=X and vacancies $v$ in the crystal lattice of pores, with
the ``host'' atoms  $h$ being vacancies $v$. The total
configurational Hamiltonian (\ref{H^t}) here includes \hbox{X-X}
interactions between atoms X, but not X-$v$ and $v$-$v$
interactions. Therefore, only $V_{ij}^{\rm XX}$ terms are nonzero in
formulas (\ref{H^t})-(\ref{v_il}):
\begin{equation}
V_{ij}^{\alpha\alpha}=V_{ij}^{\rm XX}\neq 0, \qquad V_{ij}^{\alpha
h}=V_{ij}^{hh}=0.\label{V^XX}
\end{equation}
The only meaningful index $\alpha$=X is usually skipped below, for
example: \, $c_i^{\rm X}=c_i$,\, $V_{ij}^{\rm XX}=v_{ij}$, thus the
effective Hamiltonian (\ref{H_eff}) takes the form:
\begin{equation}
H_{eff}=-\sum_{i}\lambda_in_i+H_{int},\qquad
H_{int}=\frac{1}{2}\sum_{ij}v_{ij}n_in_j.\label{H_eff-X}
\end{equation}
The mean occupation  $c_i=\langle n_i\rangle$ of pore $i$ by an atom
X is related to  the local chemical concentration $x_i$ by the
relation depending on the geometry of pores \cite{Agren-82c}, e. g.
$c$=$x/(1-x)$ for a uniform austenite structure MeX$_c$ with the FCC
lattice of octo-pores.

An important principal feature differing the diffusional kinetics in
the interstitial Me-X (i. e., substitution X-$v$) alloys from that
in the substituton A-B-$v$ alloys is the validity for Me-X alloys of
relation (\ref{h=0}), that is, the absence of interaction
renormalization effects. It follows, first, from the physical
considerations presented after Eqs. (\ref{h=0}) and (\ref{QGD}) and,
second, from the thorough  analysis of interaction renormalization
effects for A-B-$v$ alloys made by Nastar et al
\cite{Nastar-00,Nastar-11}. These authors found these effects to be
described by the terms antisymmetric with respect to interchanging A
and B atoms, $(h_{ij}^{\rm AB}-h_{ij}^{\rm BA})$, which vanish in a
binary X-$v$ alloy where A=B=X. Therefore, the diffusional kinetics
in Me-X alloys can be described by the quasi-equilibrium relations
(\ref{h=0}) and (\ref{QGD}).

For a uniform Me-X alloy, the site chemical potential
$\lambda_i$=$\lambda$ in (\ref{H_eff-X}) coincides with the
thermodynamic chemical potential $\mu_{\rm X}$, unlike substitution
alloys where the analogous quantity $\lambda_{\alpha}$, as mentioned
in Sec. 2.1, is equal to the difference
\hbox{$(\mu_{\alpha}-\mu_h$)}. To show it, we generalize Eqs.
(21)-(24) and (40)-(43) of Ref. \cite{VZhKh-10} to the case of
interstitial alloys Me-X. The quasi-equilibrium Gibbs distribution
(\ref{QGD}) and the generalized grand canonical potential
\,$\Omega_g$\, for the effective Hamiltonian (\ref{H_eff-X}) have
the form
\begin{eqnarray}
&&P=\exp \Big[\beta(\Omega_g+\sum_{\alpha i}\lambda_{\alpha
i}n_{\alpha i}-H_{int})\Big],\label{P_X}\\
&&\Omega_g=-T\ln \sum_{\{n_{\alpha i}\}} \exp
\Big[\beta\Big(\sum_{\alpha i}\lambda_{\alpha i}n_{\alpha
i}-H_{int}\Big)\Big],\label{Omega_g}
\end{eqnarray}
while the mean occupation $c_i$ is related to
\,$\Omega_g\{\lambda_{\alpha i}\}$\, by the formula obtained by the
differentiation of equality (\ref{Omega_g}):\,
\begin{equation}
c_i=\langle n_i\rangle
=-\partial\Omega_g/\partial\lambda_i.\label{c_i-Omega_g}
\end{equation}
Therefore, if we define the generalized free energy \,$F$\,
 by the equality
\begin{equation}
F=\Omega_g+\sum_{i}\lambda_{i}c_{i}\,,\label{F_tot}
\end{equation}
then site chemical potential $\lambda_{i}$ is related to \,$F$\, by
the relations generalizing a similar relation for a uniform alloy:
\begin{equation}
\lambda_{i}=\partial F/\partial c_i.\label{lambda_i-F}
\end{equation}
To relate $\lambda_{i}$ and \,$\Omega_g$\, in Eqs.
(\ref{P_X})-(\ref{lambda_i-F}) to the thermodynamic chemical
potentials, we consider the case of a uniform alloy Me-X when  $c_i$
and $\lambda_i$ in Eqs. (\ref{P_X}-(\ref{lambda_i-F}) are
independent of $i$:\, \,$c_{ i}=c$,\, $\lambda_i$=$\lambda$. For
definiteness, we discuss the austenite structure for which the total
number of interstitial sites (octo-pores) is equal to  the total
number $N_{\rm Me}$ of Me atoms. Then instead of the total
thermodynamic potentials, \,$\Omega_g$\,  and \,$F$,\, it is
convenient to consider the analogous quantities per one atom Me,
\,$\Omega$\, and \,$f$:
\begin{equation}
\Omega=\Omega_g/N_{\rm Me},\qquad f=F/N_{\rm Me}=\Omega+\lambda
c,\qquad c=N_{\rm X}/N_{\rm Me}. \label{Omega-f}
\end{equation}
Here  $N_{\rm X}$ is the total number of atoms X, thus $c$ is the
mean occupation of an interstitial site, and, according to Eq.
(\ref{lambda_i-F}):
\begin{equation}
\lambda=\partial f/\partial c.\label{lambda-f}
\end{equation}
The quantities \,$\Omega$\, and \,$\lambda$\,  in Eqs.
(\ref{Omega-f}) and (\ref{lambda-f}) are simply related to the
partial chemical potentials, \,$\mu_{\rm X}$ and \,$\mu_{\rm Me}$,\,
defined by the thermodynamic relations:
\begin{equation}
\mu_{\rm X}=\partial F/\partial N_{\rm X},\quad\mu_{\rm Me}=\partial
F/\partial N_{\rm Me}. \label{mu_X,Me}
\end{equation}
Substituting relations  (\ref{Omega-f}) for  \,$c$\, and \,$F=N_{\rm
Me}f(c)$\, in Eqs. (\ref{mu_X,Me}) and taking into account Eq.
(\ref{lambda-f}), we obtain:
\begin{equation}
\lambda=\mu_{\rm X}, \qquad\Omega=\mu_{\rm
Me}.\label{lambda,Omega-mu_X,Me}
\end{equation}
Thus, the quantity \,$\lambda$\, or  \,$\Omega$\, in Eqs.
(\ref{Omega-f}) and (\ref{lambda-f}) has the meaning of the chemical
potential of atoms X or atoms Me, respectively.

Kinetic equation describing diffusion of atoms X in an interstitial
alloy Me-X (treated as a binary substitution alloy X-$v$) can be
obtained by substitution of relations (\ref{V^XX}) and
(\ref{H_eff-X}) into the first Eq. (\ref{c_alpha-v-dot}):
\begin{equation}
dc_i/dt=\sum_{j(i)}\gamma_{ij}b_{ij}^{\rm
X}[\exp(\beta\lambda_j)-\exp(\beta\lambda_i)].\label{c_i-dot}
\end{equation}
Here we took into account that for a host atom $h=v$, the site
chemical potential $\lambda_i^h$ in  (\ref{c_alpha-v-dot}) should be
formally put zero, as illustrated by Eqs. (\ref{n^h,v-term}) and
(\ref{c_alpha-v-dot}). The jump probability \,$\gamma_{ij}$\,  and
the correlator \,$b_{ij}^{\rm X}$\, in (\ref{c_i-dot}) are defined
by the relations analogous to (\ref{gamma^pv}) and (\ref{b_ij^p}):
\begin{eqnarray}
&&\gamma_{ij}=\omega_{ij}^{eff}\exp\,(-\beta
E_{ac}^{ij}),\label{gamma_ij}\\
&&b_{ij}^{\rm X}=\Big\langle (1-n_i)(1-n_j)\exp
\Big(-\beta\sum_k\Delta^{ij}_kn_k\Big)\Big\rangle.
 \label{b_ij-X}
\end{eqnarray}
The pre-exponent $\omega_{ij}^{eff}$ in (\ref{gamma_ij}) is
determined by Eq. (\ref{omega_pv}) with replacing $p\to$X, while the
activation energy $E_{ac}^{ij}$ is reduced to the  term
$E^p_h=E^{\rm X}_v$ in (\ref{E_sp}), unlike the more complex
expression (\ref{E_ac-hv}) in a substitutional alloy. Index ``$ij$''
at the quantities $\omega_{ij}^{eff}$ and $E_{ac}^{ij}$ in
(\ref{gamma_ij}) allows for a possible nonuniformity of an alloy;
for a uniform alloy this index can be omitted. The quantity
$\Delta^{ij}_k$ in (\ref{b_ij-X}) is the analogue of
$\Delta^{p}_{ij}$ in Eqs. (\ref{E_sp}) and (\ref{E_h^p}); it
describes the change of the saddle point energy $E^{SP}_{{\rm
X}i,vj}$ for an inter-site X-atom jump $i$$\to$$ j$ due to the
presence of another atom X at site $k$.

The kinetic equation (\ref{c_i-dot}) can also be written in a form
analogous to Eq. (\ref{c_alpha-beta-sinh}):
\begin{equation}
dc_i/dt=\sum_{j(i)}2M_{ij}\sinh\big[\beta
(\lambda_j-\lambda_i)/2\big]\label{c_i-dot_M_ij}
\end{equation}
where the generalized mobility $M_{ij}$, according to
(\ref{c_i-dot}),  is determined by the relation:
\begin{equation}
M_{ij}=\gamma_{ij}b_{ij}^{\rm
X}\exp\big[\beta\big(\lambda_i+\lambda_j\big)/2].\label{M_ij}
\end{equation}

In usual diffusion problems, the space dependence of functions
$c_i=c({\bf r}_i)$, $\lambda_i=\lambda (c_i)$, and $b_{ij}^{\rm
X}=b_{ij}^{\rm X}(c_i,c_j)$ in Eqs. (\ref{c_i-dot})-(\ref{M_ij}) is
supposed to be smooth. Therefore, variations of these functions
under replacing $c_i\to c_j$\,  (or ${\bf r}_{i}$$\to$${\bf
r}_{j}$$={\bf r}_{i}+{\bf r}_{ji}$ where \,${\bf r}_{ji}={\bf
r}_{j}-{\bf r}_{i}$\, is  the inter-pore distance) are small.  Then
the kinetic equation (\ref{c_i-dot}) or (\ref{c_i-dot_M_ij}) can be
expanded in powers of ${\bf r}_{ji}\nabla c$. It yields the
continuous version of this kinetic equation having the form of
the``continuity equation'' for the flow ${\bf j}$ of atoms X:

\begin{equation}
\partial c/\partial t+{\rm div}{\bf j}=0,\quad
j_{\alpha}=-\sum_{\nu}D_{\alpha\nu}(c)\nabla_{\nu}c\,.
\label{c_i-dot-div_j}
\end{equation}
Here \,$\alpha$\, and $\nu$ are Cartesian indices, and the
diffusivity $D_{\alpha\nu}$ is determined by the following
expression:
\begin{eqnarray}
&&D_{\alpha\nu}(c)=\Gamma_{\alpha\nu}b_{\rm
X}\partial a_{\rm X}/\partial c,\label{D_alpha-nu}\\
&& \Gamma_{\alpha\nu}=
\frac{1}{2}\sum_{j(i)}\gamma_{ij}r_{ij}^{\alpha}r_{ij}^{\nu}\label{Gamma}
\label{varphi-c}
\end{eqnarray}
where  \,$a_{\rm X}=a_{\rm X}(c,T)=\exp\,[\beta\lambda (c,T)]$\, is
the thermodynamic activity of X atoms, and \,$b_{\rm X}=b_{\rm
X}(c,T)$\, is the correlator \,$b_{ij}^{\rm X}$\, in Eq.
(\ref{b_ij-X}) at \,$c_i=c_j=c$.

The site chemical potential $\lambda_i$ in a binary alloy can be
written as
\begin{equation}
\lambda_i=\lambda_i^{id}+\lambda_i^{int}
 \label{lambda_int}
\end{equation}
where \,$\lambda_i^{id}=T\ln[c_i/(1-c_i)]$\, corresponds to the
ideal solution, and $\lambda_i^{int}$ describes interaction effects,
see, e. g., (\ref{lambda_PCA}).  Therefore, for the ideal solution
for which both $\lambda_i^{int}$ in  (\ref{lambda_int}) and
$\Delta^{ij}_l$ in (\ref{b_ij-X}) are zero, we have: \,$a_{\rm
X}=c/(1+c)$,\, \  \,$b_{\rm X}=(1-c)^2$,\,  \,$b_{\rm X}\partial
a_{\rm X}/\partial c=1$,\, and Eq. (\ref{c_i-dot-div_j}) takes the
form of a simple linear diffusion equation:
\begin{equation}
\partial c/\partial t=\sum_{\alpha\nu}D_{\alpha\nu}^{id}\nabla^2_{\alpha\nu}c
\label{c_-dot-D}
\end{equation}
with the concentration-independent diffusivity
$D_{\alpha\nu}^{id}$\, equal to $\Gamma_{\alpha\nu}$ in
(\ref{Gamma}). However, when X-X interactions \,$v_{ij}$\, and
\,$\Delta^{ij}_l$\, are significant, the kinetic equation
(\ref{c_i-dot-div_j}) is nonlinear, and the diffusivity \,$\bf D$\,
in (\ref{D_alpha-nu}) should vary with the local concentration
$c=c^{\rm X}({\bf r})$.

For a uniform cubic alloy, such as austenite, tensor
$\Gamma_{\alpha\nu}$ is reduced to a scalar
\,$\delta_{\alpha\nu}\gamma a^2$\, where $\gamma$ is given by Eq.
(\ref{gamma_ij}) (with omission of index \,$ij$)\, and $a$ is the
FCC iron ($\gamma$-iron) lattice constant, hence
\,$D_{\alpha\nu}=\delta_{\alpha\nu}D$.\, Then the diffusivity
\,$D$,\, according to  (\ref{D_alpha-nu}), can be written as
\begin{equation}
D=\gamma a^2b_{\rm X}\,\partial a_{\rm X}/\partial c \label{D_X-gen}
\end{equation}
where \,$\gamma$\, is defined by Eqs. (\ref{gamma_ij}) and
(\ref{omega_pv}) with the appropriate change of indices:
\begin{equation}
\gamma=\omega_{{\rm X}}^{eff}\exp(-\beta E_{ac}),\quad \omega_{{\rm
X}}^{eff}=\omega_{{\rm X}}\exp\big(\Delta S_{{\rm
X}}^{SP}\big).\label{gamma_X}
\end{equation}
For a uniform alloy with $c_i=c$, the PCA expression
(\ref{lambda_PCA}) for  the local chemical potential $\lambda (c)$
is simplified \cite{VKh-1}:
\begin{eqnarray}
&&\lambda(c,T)=T\sum_nz_n\ln (1-g_nc),\qquad
g_n=2f_n/(R_n+1+2cf_n),\nonumber\\
&& f_n=\exp\,(-\beta v_n)-1,\qquad
R_n=[1+4c(1-c)f_n]^{1/2}\label{lamba_PCA-unif}
\end{eqnarray}
where $z_n$ is the coordination number, and $v_n$ is the
configurational interaction for the $n$th coordination sphere. In
the case of weak interaction, \,$\beta v_n\ll1$,\, Eq.
(\ref{lamba_PCA-unif}) is reduced to the MFA expression:
\,$\lambda^{\rm MFA}=(\sum_nz_nv_n)\,c$.\, However, for the
realistic values of interactions $v_n$, such as those presented in
Table 1 below, using the MFA can lead to significant errors
\cite{VKh-1}.

Correlator $b_{\rm X}(c,T)$ for a uniform alloy, according to Eqs.
(\ref{b_ij-X}) and (\ref{tilde_b_ij^p}), can be written as
\begin{equation}
b_{\rm X}(c,T)=(1-c)^2 \exp\Big[\sum_{n=1}
z_n^{sp}\ln(1+f_n^{sp}c)\Big],\qquad
f_n^{sp}=\exp\,(-\beta\Delta_n)-1 \label{b^X}
\end{equation}
where  $z_n^{sp}$ is the coordination number, and $\Delta_n$ is the
saddle-point interaction for the $n$th coordination sphere of the
saddle point considered. If these interactions are weak: $\beta
\Delta_n\ll 1$, Eq. (\ref{b^X}) takes its MFA form:
\begin{equation}
b_{\rm X}(c,T)=(1-c)^2 \exp\Big(-\beta c\sum_{n=1}
z_n^{sp}\Delta_n\Big) \label{b^X_MFA}
\end{equation}
However, for the realistic saddle-point interactions $\Delta_n$,
such as those shown in Table 2 below,  using the MFA can lead to
significant errors, just as for $\lambda$ in Eq.
(\ref{lamba_PCA-unif}).

The microscopic relation (\ref{D_X-gen}) can be compared to various
phenomenological models for diffusivity
\cite{Agren-82a}-\cite{Bhadeshia-81}. It can provide, in particular,
the statistical expression for the phenomenological mobility $M_{\rm
CVa}$ introduced by {\AA}gren in his discussion of diffusion of
carbon in austenite \cite{Agren-82a}-\cite{Agren-86}. Comparing Eq.
(\ref{D_X-gen}) to the definition of $M_{\rm CVa}$  by equation (9)
in \cite{Agren-82c}, we find:
\begin{equation}
M_{\rm CVa}=\gamma a^2b_{\rm X}\,a_{\rm C}\Big/[c(1-c)V_mT]
\label{M_c-Va}
\end{equation}
where \,$V_m$\, is the volume per atom Me. In more detail, the
microscopic and  phenomenological descriptions of diffusion of
carbon in austenite are compared in Sec. 4.

Let us now consider a multicomponent interstitial alloy
(Me$_1$Me$_2$...)-X with several species atoms $p$ in the metal
sublattice, such as an Fe-Mn-C alloy. The interstitial sites will be
marked by indices $i$, $j$ and $k$, while the sites in the metal
sublattice, by indices $l$, $m$ and $n$. The total configurational
Hamiltonian can be written in the form generalizing Eq. (\ref{H^t}):
\begin{equation}
H^t={1\over 2}\sum_{ij}v_{ij}n_in_j+
+\sum_{p,il}V_{il}^{p}n_in_l^{p}+ {1\over
2}\sum_{pq,lm}V_{lm}^{pq}n_l^{p}n_m^{q} \label{H^t-X}
\end{equation}
where we again skip index X for an interstitial atom putting
$n_i^{\rm X}=n_i$,\, $V_{il}^{{\rm X}p}=V_{il}^{p}$, \,$V_{ij}^{\rm
XX}=v_{ij}$. As above, we discuss diffusion of only interstitial but
not metal atoms, and the presence of vacancies in the metal
sublattice is neglected.  Therefore, occupation operators \,$n_l^h$
for host metal $h$ can be expressed via those for the minority
metals $\alpha$ similarly to Eq. (\ref{n^h}):
\,$n_l^h=(1-\sum_{\alpha}n_l^{\alpha})$.\, The effective Hamiltonian
for statistical averaging, instead of (\ref{H_eff}), takes the form:
\begin{eqnarray}
&H_{eff}=&-\sum_{i}\lambda_in_i
-\sum_{\alpha,l}\lambda_l^{\alpha}n_l^{\alpha}\nonumber\\
&&+{1\over 2}\sum_{ij}v_{ij}n_in_j
+\sum_{\alpha,il}v_{il}^{\alpha}n_in_l^{\alpha}+ {1\over
2}\sum_{\alpha\beta,lm}v_{lm}^{\alpha\beta}n_l^{\alpha}n_m^{\beta}\label{H_eff-X-ligands}
\end{eqnarray}
where $v_{il}^{\alpha}=(V_{il}^{\alpha}-V_{il}^{h})$, while
$v_{lm}^{\alpha\beta}$ is related to $V^{pq}_{lm}$ in (\ref{H^t-X})
similarly to Eq. (\ref{v_il}).

Equations describing diffusion  of atoms X can be again derived
using Eqs. (\ref{dP/dt})-(\ref{n^h,v-term}) with appropriate
generalizations and simplifications. In particular, the first
equation (\ref{dc_v-dt}) here takes the form
\begin{equation}
dc_i/dt=\sum_{\{n_k,n_l^{\alpha}\}}\sum_{j(i)} \omega_{\rm
X}^{eff}\Big[(1-n_i)n_j\exp\big(\beta \hat E_{ij}^{in}-\beta\hat
E_{ij}^{SP}\big)-\{i\to j\}\Big]\exp[\beta
\big(\Omega-H_{eff}\big)].\label{dc_i-dt-ligand}
\end{equation}
Here the saddle-point energy $\hat E_{ij}^{SP}$, instead of Eqs.
(\ref{E_sp}) and (\ref{E_h^p}), is given by the expression
\begin{equation}
\hat E_{ij}^{SP}=E_h+\sum_k\Delta^{ij}_kn_k+\sum_{\alpha
l}\Delta^{ij}_{\alpha l}n^{\alpha}_l \label{E-sp-ligand}
\end{equation}
where $\Delta^{ij}_k$ and $\Delta^{ij}_{\alpha l}$ are analogues of
$\Delta^{ij}_k$  in  (\ref{b_ij-X}). Using relations (\ref{E^in})
and (\ref{n^h,v-term}), we again reduce Eq. (\ref{dc_i-dt-ligand})
to the form  (\ref{c_i-dot}). However, the activation energy
$E_{ac}$  in  (\ref{gamma_ij}) and the correlator $b_{ij}^{\rm X}$
in (\ref{b_ij-X}) now are defined as
\begin{eqnarray}
&&E_{ac}=E_h-\sum_lV_{il}^h,\nonumber\\
&&b_{ij}^{\rm X}=\Big\langle (1-n_i)(1-n_j)\exp
\Big(-\beta\sum_k\Delta^{ij}_kn_k-\beta\sum_{\alpha
l}\Delta^{ij}_{\alpha l}n^{\alpha}_l\Big)\Big\rangle,
\label{b_ij-X-ligands}
\end{eqnarray}
and Eqs. (\ref{dc_i-dt-ligand}) and (\ref{b_ij-X-ligands}) now
include the statistical averaging over various distributions of
${\alpha}$-species atoms in the metal sublattice.

\section{CALCULATIONS OF DIFFUSIVITY AND ACTIVITY OF
CARBON IN AUSTENITE FOR SIMPLE MODELS OF CARBON-CARBON INTERACTIONS}

To calculate the diffusivity \,$D$\, according to the microscopic
expression (\ref{D_X-gen}), we should use some theoretical model of
X-X interactions in an alloy, for both the configurational
interactions $v_n$ in (\ref{lamba_PCA-unif}) which determine the
chemical potential $\lambda$, and for the saddle-point interactions
$\Delta_n$ in (\ref{b^X}) which  determine the correlator $b_{\rm
X}$. For substitution Fe-Cu alloys, such first-principle model for
both $v_n$ and $\Delta_n$ has been developed by SF \cite{SF-07}, and
simulations of precipitation in  Fe-Cu alloys based on this model
confirmed its adequacy and reliability \cite{SF-07}. For
interactions of carbon in austenite, reliable first-principle
calculations are still absent due to the well-known difficulties of
taking into account magnetic interactions in $\gamma$-iron
\cite{Jiang-03}. However, some simplified model of configurational
interactions $v_n$ in austenite has been suggested by Blanter
\cite{Blanter-99}, and his estimates of  these interactions are
presented in Table 1 as $v_n^B$. Below we use this model and some
its extensions to investigate  the concentration and temperature
dependences of diffusivity $D$ which follow from the microscopic
expression (\ref{D_X-gen}).

Blanter used  the model of purely deformational configurational
interactions with the nearest-neighbor Kanzaki forces for all
constants $v_n^B$ except the first one. The nearest-neighbor
constant $v_1^B$ (which can not be adequately described by the
deformational model due to the strong ``chemical'' repulsion at
short C-C distances) was treated as a free parameter which was
estimated from the fit of the carbon activity in austenite with
respect to graphite, $a_{\rm C}^{\gamma-gr}$, calculated with these
$v_n^B$ to the experimental values. The quantity $a_{\rm
C}^{\gamma-gr}$ is related to the ``configurational'' activity
$a_{\rm C}=\exp \,(\beta\lambda_{\rm C})$, where $\lambda_{\rm
C}=\lambda$ is the chemical potential of carbon in austenite
discussed in Sec. 3, by the thermodynamic relation
\cite{Blanter-99}:
\begin{equation}
a_{\rm C}^{\gamma-gr}=a_{\rm C}\exp(\beta\Delta G^{\gamma-gr}_{\rm
C}) \label{a_C}
\end{equation}
where $\Delta G^{\gamma-gr}_{\rm C}=\Delta G^{\gamma-gr}_{\rm C}(T)$
is the difference between the thermodynamic potentials per carbon
atom in a pure $\gamma$-iron and in graphite. The fit to
experimental $a_{\rm C}^{\gamma-gr}(c,T)$ obtained with the use for
both Monte Carlo \cite{Blanter-99}, and the PCA \cite{VKh-1}
calculations of $a_{\rm C}(c,T)$ the $v_n^B$ values, and for $\Delta
G^{\gamma-gr}_{\rm C}(T)$, some experimental estimates, seemed to be
quite satisfactory. It may imply that the simple model of Blanter
\cite{Blanter-99}  can serve as a basis for realistic descriptions
of C-C interactions in austenite.

\vskip6mm

\vbox{ \noindent{\bf Table 1.} Configurational interactions $v_n$
(in kelvin) of carbon atoms in austenite

\vskip3mm {\begin{tabular}{|c|ccccccccccc|} \hline &&&&&&&&&&&\cr
$n$&1&2&3&4&5&6&7&8&9&10&11\cr\hline  &&&&&&&&&&&\cr $2{\bf R}_n/a$
&110&200&211&220&310&222&321&400&330&411&420\\%\hline &&&&&&&&&&&\cr
$R_n/R_1$ &1&1.41&1.73&2&2.24&$2.45$&$2.65$&2.83&3&3&3.16\\
 $z_n$ &12&6&24&12&24&8&48&6&12&24&24\\\hline &&&&&&&&&&&\cr
$v_n^B$, Blanter
\cite{Blanter-99}&1334&1961&-487&46&46&267&-23&-139&58&-12&-23\\%\hline
$v_n$, this
work&1400&1180&-322&46&46&267&-23&-139&58&-12&-23\\\hline
\end{tabular}}}

\vskip6mm

In the present work, the configurational interactions  $v_n$ have
been estimated using a similar  approach. However, in the fit to
experimental values $a_{\rm C}^{\gamma-gr}(c,T)$ we used for the
function $\Delta G^{\gamma-gr}_{\rm C}$ in (\ref{a_C}) the
interpolation of experimental data  suggested by \AA gren
\cite{Agren-79}:
\begin{equation}
\Delta G^{\gamma-gr}_{\rm C}=5550\,{\rm K}-2.31\,T
\label{Delta_G-gamma-gr}
\end{equation}
rather than that used in \cite{VKh-1}, and we also varied not only
$v_1$ but also two next constants, $v_2$ and  $v_3$. The  $v_n$
values obtained are presented in the last line of Table 1.
Variations of our $v_2$ and $v_3$ with respect to their ``purely
deformational'' values $v_2^B$ and $v_3^B$ lie certainly within the
real accuracy of the original Blanter model as, first, this model
disregards ``chemical'' contributions to $v_2$ and $v_3$ which can
be quite notable (which is illustrated, in particular, by comparison
of results of calculations of C-C interactions in ferrite based on
$ab$-$initio$ \cite{Domain-04} and purely deformational
\cite{Blanter-78} approaches \cite{VKh-5}) and, second, it neglects
both possible contributions of not-nearest Kanzaki forces
\cite{VKh-1} and a probable variation of phonon spectra with
temperature (which was not measured in $\gamma$-iron but is very
pronounced in the BCC iron \cite{Satya-85,Neuhaus-97}). In Figs. 1
and 2 we present the carbon activity $a_{\rm C}^{\gamma-gr}(x_{\rm
C},T)$  and the equilibrium phase diagram ferrite-austenite
calculated using the PCA expression (\ref{lamba_PCA-unif}) for
$\lambda$ with our $v_n$ from Table 1, together with experimental
data and the results of calculations based on the phenomenological
model by \AA gren \cite{Agren-79}.
\begin{figure}
%\begin{center}
%\psfig{file=dv98_2.eps,scale=0.38, angle=0}\hspace{5mm}
%\end{center}
\caption{Dependence of carbon activity $a_{\rm C}^{\gamma-gr}$ in
austenite with respect to graphite on the carbon concentration
$x_{\rm C}=c/(1+c)$ for various temperatures $T$.  Dots correspond
to experimental data presented in \cite{Blanter-99}. Solid curves
are calculated using the PCA expression (\ref{lamba_PCA-unif}) for
$\lambda$ with the interaction constants $v_n$ from Table 1. Dashed
curves are calculated using the phenomenological description of C-C
interactions employed by \AA gren \cite{Agren-79}.\label{a_C-c,T}}
\end{figure}
\begin{figure}
%\begin{center}
%\psfig{file=dv98_2.eps,scale=0.38, angle=0}\hspace{5mm}
%\end{center}
\caption{Fe-C phase diagram. Dots correspond to experimental phase
boundaries. Solid curves show ferrite-austenite phase boundaries
calculated using Eq. (\ref{lamba_PCA-unif}) with $v_n$ from Table 1.
Dashed curves correspond to the phenomenological calculations by \AA
gren \cite{Agren-79}.\label{Fe-C-phase-diagram}}
\end{figure}

\begin{figure}
%\begin{center}
%\psfig{file=dv98_2.eps,scale=0.38, angle=0}\hspace{5mm}
%\end{center}
\caption{Illustration of our method of estimates of saddle-point
interactions $\Delta_n$ using interpolation of configurational
interactions $v_n=v(R_n)$. Circles show values of $v_n$, and
triangles, values of $\Delta_n^{(0)}$ obtained as described in the
text. \label{Delta_n-v_n-method}}
\end{figure}

\begin{figure}
%\begin{center}
%\psfig{file=dv98_2.eps,scale=0.38, angle=0}\hspace{5mm}
%\end{center}
\caption{Values of carbon-carbon interactions used in the present
work. Open circles: configurational interactions  $v_n=v(R_n)$;
black triangles: saddle-point interactions $\Delta_n=
\Delta(R_n^{sp})$. Dashed lines connect the neighboring $\Delta_n$
values to guide the eye. \label{v_n-Delta_n-final}}
\end{figure}

\vskip3mm
\vbox{
\noindent{\bf Table 2.} Saddle-point interactions $\Delta_n=
\Delta({\bf R}_{n}^{sp})$ (in kelvin) of carbon atoms in austenite
for vectors ${\bf R}_{n}^{sp}=({\bf R}_{n}'-{\bf R}_{sp})$ where
${\bf R}_{sp}$ is the saddle-point position of carbon atom

\vskip3mm {\begin{tabular}{|c|cccccccccc|} \hline &&&&&&&&&&\cr
$n$&1&2&3&4&5&6&7&8&9&10\cr\hline  &&&&&&&&&&\cr
$R_{n}^{sp}/R_1$ &0.87&1.12&1.32&1.5&1.66&$1.80$&$1.94$&2.06&2.18&2.29\\
 $z_{n}^{sp}$ &4&4&8&6&4&12&8&8&12&8\cr\hline &&&&&&&&&&\cr
$\Delta_n$&1470&1336&1228&229&-924&-929&-543&133&133&286\\\hline
\end{tabular}}

\vskip3mm {\begin{tabular}{|c|ccccccccc|} \hline &&&&&&&&&\cr
$n$&11&12&13&14&15&16&17&18&19\cr\hline &&&&&&&&&\cr
$R_{n}^{sp}/R_1$ &2.40&2.50&2.60&2.70&2.78&2.87&2.96&3.04&3.12\\
 $z_{n}^{sp}$ &8&14&16&4&16&16&8&20&8\cr\hline &&&&&&&&&\cr
$\Delta_n$&622&564&144&-160&-310&-269&34&-43&-59\cr \hline
\end{tabular}}}

 \vskip6mm
Let us now discuss the saddle-point interactions $\Delta_n=
\Delta({\bf R}_{n}^{sp})$ where ${\bf R}_{n}^{sp}=({\bf R}'_{n}-{\bf
R}_{sp})$, and ${\bf R}_{sp}$ is the saddle-point position of carbon
atom. In the second and the third lines of Table 2 we show the first
19 distances ${R}_{n}^{sp}=|{\bf R}_{n}^{sp}|$ and the coordination
numbers $z_n^{sp}$ that correspond to these ${\bf R}_{n}^{sp}$. To
illustrate the distribution of vectors ${\bf R}_{n}^{sp}$ in the FCC
lattice, below we present the values of components of lattice
vectors ${\bf R}'_{n}=({\bf R}_{n}^{sp}+{\bf R}_{sp})$ (in $a/2$
units) for the first eight coordination spheres of the point ${\bf
R}_{sp}$=(0.5,0.5,0):
\begin{eqnarray}
&&\hskip-6mm{\bf R}_{1}'=(0,1,\pm 1), (1,0,\pm 1);\qquad {\bf
R}_{2}'=(0,2,0), (2,0,0), (1,\bar 1,0), (\bar 1,1,0);
\nonumber\\
&&\hskip-6mm{\bf R}_{3}'=(2,1,\pm 1), (1,2,\pm 1), (0,\bar 1,\pm 1),
(\bar
1,0,\pm 1);\nonumber\\
&&\hskip-6mm{\bf R}_{4}'=(2,2,0), (\bar 1,\bar 1,0), (0,0,\pm 2),
(1,1,\pm
2); \qquad {\bf R}_{5}'=(2,\bar 1,\pm 1), (\bar 1,2,\pm 1);\nonumber\\
&&\hskip-6mm {\bf R}_{6}'=(0,\bar 2,0), (\bar 2,0,0), (1,3,0),
(3,1,0),(0,2,\pm 2), (2,0,\pm 2), (1,\bar 1,\pm 2), (\bar 1,1,\pm
2);(\bar 1,1,\pm 3);\nonumber\\
&&\hskip-6mm{\bf R}_{7}'=(0,3,\pm 1), (3,0,\pm 1), (\bar
2,1,\pm 1), (1,\bar 2,\pm 1), (3,0,\pm 1);\nonumber\\
&&\hskip-6mm{\bf R}_{8}'=(2,\bar 2,0), (\bar 2,2,0), (3,\bar 1,0),
(\bar 1,3,0), (0,1,\pm 3), (\bar 1,\bar 1,\pm 1). \label{R_n^sp}
%&&\hskip-6mm{\bf R}_{9}^{sp}=(0,1,\pm 3), (1,0,\pm 3);\qquad {\bf
%R}_{10}^{sp}=(0,\bar 2,\pm 2), (\bar 2,0,\pm 2), (1,3,\pm 2),
%(3,1,\pm 2); \nonumber\\&&\hskip-6mm{\bf R}_{11}^{sp}=(0,3,\pm 3), (3,0,\pm 3),(0,\bar 2,\pm
%3), (\bar2,0,\pm 3), (1,\bar 1,\pm 3);\nonumber\\&&\hskip-6mm {\bf R}_{12}^{sp}
%=(0,4,0), (4,0,0), (1,\bar 3,0), (\bar 3,1,0)\nonumber\\.
\end{eqnarray}

Models for estimation of the saddle-point interactions $\Delta_n$
can be constructed similarly to those for the configurational
interactions $v_n$. The long-ranged deformational contributions to
$\Delta_n$ can be calculated using the general expression for
deformational interactions with the arbitrary Kanzaki forces
discussed in \cite{VKh-1}, while the short-range chemical
contributions can be estimated treating several first  $\Delta_n$ as
adjustable parameters, as it was made for the $v_n$. However, in
this work we restrict ourselves by illustrative estimates of
$\Delta_n$ based on some interpolations between $v_n$ values in
Table 1 and several simple assumptions. First, we assume that both
the chemical and the deformational contributions to $\Delta_n$
depend only on the distance ${R}_{n}^{sp}=|{\bf R}_{n}^{sp}|$ and
vary with ${R}_{n}^{sp}$ smoothly. Second, we assume that for short
distances $R_n^{sp}<R_2$=1.41$R_1$, the $\Delta_n$ values are mainly
determined by the chemical contributions, while for longer
$R_n^{sp}>R_2$, these values are mainly determined by  the
deformational contributions. Third, we assume that the dependence of
the configurational interactions $v_n$ on $R_n$ is similar to the
dependence of the saddle-point interactions $\Delta_n$  on
${R}_{n}^{sp}$, for both chemical and deformational contributions.
Then the ``chemical'' interactions $\Delta_1$, $\Delta_2$ and
$\Delta_3$ can be estimated using the linear interpolation between
$v_1$ and $v_2$ values, as shown in Figs. 3 and 4. For the
``deformational'' $\Delta_n$ with $n\geq 4$ or $R_n^{sp}>R_2$, the
analogous estimate of $\Delta_n$ includes the following two steps:

(A) Interpolation of dependence $v(R)$ using $v_n$ values in Table
1, which yields ``preliminary'' values $\Delta_n^{(0)}$ shown in
Fig. 3.

(B) Scaling of these  $\Delta_n^{(0)}$ by some factor $\alpha$,
\begin{equation}
\Delta_n=\alpha \Delta_n^{(0)}, \label{Delta-alpha}
\end{equation}
with the value  $\alpha$ determined from the fit of the diffusivity
$D$ calculated according to Eqs. (\ref{D_X-gen})-(\ref{b^X}) to the
experimental data about the diffusivity of carbon in austenite.

The first step (A) is illustrated by Fig. 3. This figure shows that
to obtain an adequate interpolation $v(R)$, the regions of long and
``intermediate'' distances $R$ should be treated differently. At
long distances $R>R_4$, we can use the simple linear interpolation
between neighboring $v_n$ values, while at $R_2<R<R_4$, some smooth
curve should be drawn between $v_2$, $v_3$  and $v_4$ values. For
these intermediate $R$, we interpolated $v(R)$  by a simple power
law:
\begin{eqnarray}
&&R_2<R<R_3:\qquad v(R)=C_2(R_3-R)^m,\nonumber\\
&&R_3<R<R_4:\qquad v(R)=C_4(R-R_3)^m \label{v_R-2-4}
\end{eqnarray}
where constants $C_2$ and $C_4$ are determined by the conditions:
$v(R_2)=v_2$, $v(R_4)=v_4$. For the exponent $m$ we tried two
values, 2 and 4, and the value $m=4$ was found to be more suitable
for the fit mentioned in the point (B). The resulting interpolation
$v(R)$ is shown in Fig. 3 by the dashed curve.

For the second step (B), the physical arguments in favor of the
model relation (\ref{Delta-alpha}) can be seen from the general
expression for deformational interactions given by Eq. (11) in
\cite{VKh-1}. According to this expression, the deformational
interaction $V_{ij}^d=V^d({\bf R}_i-{\bf R}_j)$ between two atoms
positioned at ${\bf R}_i$ and ${\bf R}_j$ is proportional to the
integral over wave-numbers ${\bf k}$ in the Brillouin zone of some
expression that includes the product of two appropriate Kanzaki
forces, ${\bf f}^i_{\bf k}$ and  ${\bf f}^j_{\bf k}$, while each of
these forces is proportional to the amplitude of displacements of
neighboring host (iron) atoms due to the presence of an impurity
(carbon) atom at site $i$ or $j$. Therefore, for the configurational
interactions $v_n$, the deformational contributions are proportional
to the product of two Kanzaki forces, ${\bf f}^{op}_i$ and  ${\bf
f}^{op}_j$, that describe the displacements of iron atoms induced by
a carbon atom positioned in the octo-pore. At the same time, for the
saddle-point interaction $\Delta_n$, one of these factors is
replaced by a Kanzaki force  ${\bf f}^{sp}_i$ that describes the
analogous displacements of iron atoms but induced by a carbon atom
in the saddle-point position, for which the carbon-iron distance
$R_{\rm Fe-C}$ is by $1/\sqrt{2}$ times smaller than that for a
carbon atom in an octo-pore. Therefore, this Kanzaki force ${\bf
f}^{sp}_i$ can be expected to notably exceed ${\bf f}^{op}_i$. Hence
the factor $\alpha$ in (\ref{Delta-alpha}), which qualitatively
describes the relative scale of deformational contributions to the
$v_n$ and to the $\Delta_n$ values, can notably exceed unity.
\begin{figure}
%\begin{center}
%\psfig{file=dv98_2.eps,scale=0.38, angle=0}\hspace{5mm}
%\end{center}
\caption{The diffusivity $D_{\rm C}(x_{\rm C},T)$ of carbon in
austenite. The symbols denote experimental values from
\cite{Wells-50} presented in \cite{Agren-86}. The solid lines are
calculated using Eq. (\ref{D_X-gen}) and the values $v_n$ and
$\Delta_n$ from tables 1 and 2. The dashed lines show the results of
calculations by \AA gren \cite{Agren-86} based on his
phenomenological model. \label{D_C-results}}
\end{figure}

The description of available experimental data about the diffusivity
$D_{\rm C}(x_{\rm C},T)$ \cite{Wells-50}  by our model with the
choice $\alpha=2.9$ in Eq. (\ref{Delta-alpha}) is shown in Fig. 5.
This description corresponds to the following values of the
saddle-point energy $E_{ac}$, the product $\omega_{\rm C}^{eff}a^2$,
and the frequency $\omega_{\rm C}^{eff}$ in Eqs. (\ref{D_X-gen}) and
(\ref{gamma_X}):
\begin{eqnarray}
&&E_{ac}=17700 \ {\rm K},\label{E_ac}\\
&&\omega_{\rm C}^{eff}a^2=0.225 \ {\rm cm^2/sec},\qquad \omega_{\rm
C}^{eff}=1.76\cdot 10^{14}\, {\rm sec}^{-1} \label{omega-a^2}
\end{eqnarray}
where the value $a=3.58$ \AA \, for $\gamma$-iron \cite{Kurdyumov}
is used. Let us note that the choice $\alpha=3.04$ in Eq.
(\ref{Delta-alpha}) would yield the values of $D_{\rm C}(x_{\rm
C},T)$ which virtually coincide with those obtained by \AA gren
\cite{Agren-86}. However, the choice $\alpha=2.9$ seems to better
describe the low-temperature data by Wells et al. \cite{Wells-50}
which agree with those obtained by Smith \cite{Smith-53}. The
saddle-point interactions $\Delta_n$ which correspond to
$\alpha=2.9$ are presented in Table 2 and  Fig. 4.

Both experimental and theoretical results presented in Fig. 5 show
that the diffusivity $D_{\rm C}$  sharply rises with increase of the
carbon concentration $x_{\rm C}$. In accordance with Eqs.
(\ref{E_sp}), (\ref{E_h^p}), (\ref{D_X-gen}) and (\ref{b^X}), it
seems to indicate on the presence of a significant attraction in the
saddle-point interactions $\Delta_n$, as this attraction lowers the
saddle-point energy $E^{SP}_{\rm C}$ for the inter-site jumps of
carbon atoms. In the model estimates of interactions shown in Fig.4,
it corresponds to the presence of significant negative $\Delta_n$ at
``intermediate'' carbon-carbon distances $R$ lying in the interval
$1.6R_1\lesssim R\lesssim 2R_1$. The rise $D_{\rm C}(c,T)$ with
$x_{\rm C}$ mentioned above can imply that such significant
attraction is present not only in our model estimates, but also in
the real saddle-point interactions of carbon atoms in austenite.

Let us now discuss the values of the pre-factor $\omega^{eff}_{\rm
C}$ and the ``transition state entropy'' $\Delta S^{SP}_{\rm C}$ in
(\ref{omega_pv}) which correspond to the estimate (\ref{omega-a^2}).
The attempt frequency $\omega_{pv}$ in (\ref{omega_pv}) for our case
can be estimated as the frequency $\omega_{\rm C}^{\gamma}$ of local
vibrations of carbon in austenite.  These vibrations have been
experimentally studied by Sumin et al. \cite{Sumin-90} who found:
\begin{equation}
{\omega}_{\rm C}^{\gamma}\simeq 75\, {\rm meV}= 1.14 \cdot 10^{14}\,
{\rm sec}^{-1}.\label{Delta-S_C}
\end{equation}
Note that this ${\omega}_{\rm C}^{\gamma}$  exceeds the Debye
frequency of $\gamma$-iron, $\omega_D^{\gamma}=0.43\cdot 10^{14}$
sec$^{-1}$ \cite{Stassis-87}, by about three times. Then using Eq.
(\ref{omega_pv}) with $\omega_{pv}$=$\omega_{\rm C}^{\gamma}$ and
$\omega_{pv}^{eff}$=$\omega_{\rm C}^{eff}$ from (\ref{omega-a^2}),
we obtain:
\begin{equation}
\Delta S^{SP}_{\rm C}\simeq 0.4,\qquad \bar{\omega}_{\rm
C}^{sp}\simeq 0.9\,{\omega}_{\rm C}^{\gamma}.\label{Delta-S_C}
\end{equation}
These relations show that the ``softening'' of saddle-point
frequencies ${\omega}_{\rm C}^{sp}$ with respect to ${\omega}_{\rm
C}^{\gamma}$ for carbon in austenite is rather weak (if any), unlike
Fe-Cu substitution alloys discussed in Sec. 2.1, while the
saddle-point entropy $\Delta S^{SP}_{\rm C}$ is by an order of
magnitude lower than the analogous $\Delta S^{SP}_{{\rm Cu}v}$ and
$\Delta S^{SP}_{{\rm Fe}v}$ values for the Fe-Cu alloys (as
estimated by SF \cite{SF-07}). The difference can be related (at
least, partly) to the above-mentioned inequality ${\omega}_{\rm
C}^{\gamma}\gg \omega_D^{\gamma}$ which implies that the dynamics of
carbon atoms in austenite is much faster than the iron atom
dynamics. In such conditions, an assumption of ``a local
thermodynamic equilibrium'' for the saddle-point transition state,
as well as the entropy notion for this state can be not fully
adequate and should be used with some caution.

Note that Eqs. (\ref{Delta-S_C}) correspond to the pre-factor
$\omega_{\rm C}^{eff}$ of the factor $\gamma$ in (\ref{gamma_X})
which determines the diffusivity  (\ref{D_X-gen}) in the dilute
alloy limit. Therefore, these equations have no relevance to the
illustrative estimates of carbon-carbon interactions discussed
above, but they provide some definite information about the
microscopic characteristics of diffusion of carbon in $\gamma$-iron.

Finally, let us compare the microscopic  description of
thermodynamic and diffusional characteristics of carbon in austenite
presented in this work to their phenomenological description
developed by \AA gren \cite{Agren-82c,Agren-86,Agren-79}. Both
approaches use a similar number of adjustable parameters, and the
quality of agreement between the results obtained and the
experimental data shown in Figs. 1, 2 and 5 is similar. However, the
microscopic approach seems to provide a more adequate physical
understanding of the phenomena considered. It also opens
possibilities for developments of fully first-principle descriptions
with no adjustable parameters, as demonstrated by SF \cite{SF-07}
for Fe-Cu alloys. In addition to that, the microscopic expression
(\ref{D_X-gen}) for the diffusivity seems to elucidate a number of
principal points not discussed earlier. First, it shows that the
diffusivity can be written in the form of the product of
``thermodynamic'' and ``kinetic'' (or ''saddle-point'') factors, and
the thermodynamic factors include not only the so-called
Darken-factor $\partial \ln a_{\rm X}/\partial \ln c$ usually
written \cite{Agren-82c}, but the concentration derivative of the
activity itself,  $\partial a_{\rm X}/\partial c$. Second, the
microscopic relations (\ref{D_X-gen})-(\ref{b^X}) enable us to
estimate the ``transition state entropy'' $\Delta S^{SP}$ from
experimental data, as was demonstrated for  carbon in austenite.
Third, these microscopic relations allow to relate the concentration
dependence of the activity $a_{\rm X}$ and the diffusivity $D_{\rm
X}$ to both the configurational and the saddle-point interactions
between interstitial atoms X, in particular, between carbon atoms in
austenite. Therefore, the analysis of experimental data about
$a_{\rm C}(x_{\rm C},T)$ and $D_{\rm C}(x_{\rm C},T)$ enables us to
get an idea about the  type and the scale of these interactions.

\section{CONCLUSIONS}

Let us summarize the main results of this work. The fundamental
master equation for the probability of various atomic distrubutions
in an alloy is  used to derive the basic equations of diffusional
kinetics in alloys. The microscopic parameters entering these
equations can be calculated by $ab$-$initio$ methods, as was
demonstrated by SF for iron-copper alloys \cite{SF-07}, or using
various theoretical models. For substitution alloys, the diffusional
transformation kinetics is described by the ``quasi-equilibrium''
kinetic equation (QKE) derived in Sec. 2.1. This  equation
(\ref{c_alpha-v-dot}) generalizes its earlier version presented in
\cite{KSSV-11} by taking into account possible ``interaction
renormalization'' effects which can be important for the
vacancy-mediated kinetics \cite{Nastar-00,Nastar-11}. In sec. 2.2 we
describe the calculations of local chemical potentials $\lambda_i$
and correlators $b_{ij}$ entering the QKE (\ref{c_alpha-v-dot}) with
the use of some analytical methods which combine simplicity of
calculations with a high accuracy, particularly for dilute alloys.
In Sec. 2.3 we reduce the QKE (\ref{c_alpha-v-dot}) describing the
vacancy-mediated kinetics to the kinetic equation for some
equivalent direct-atomic-exchange model which is suitable for
computer simulations.

The microscopic equations describing diffusion of interstitial atoms
X in an interstitial alloy Me-X are derived in Sec. 3. These
equations have a simple form (\ref{c_i-dot}) or
(\ref{c_i-dot_M_ij}), which enable us  to obtain the explicit
analytical expressions for the diffusivity $D=D_{\rm X}$.\, These
expressions for \,$D$\, given by Eqs. (\ref{D_alpha-nu}) or
(\ref{D_X-gen}) have a simple form of products of three factors: the
concentration derivative of the thermodynamic activity $a_{\rm X}$
of atoms X; the correlator \,$b_{\rm X}$\, given by Eqs.
(\ref{b_ij-X}) or (\ref{b^X}) which describes the influence of
interactions between atoms X on the activation barrier for the
inter-site jumps of atoms X; and the concentration-independent
factor $\gamma$ describing  the diffusivity in the dilute alloy
limit. This microscopic expression for $D$ differs notably from
those used in phenomenological treatments
\cite{Agren-82a}-\cite{Bhadeshia-81}, in particular, by the presence
of the concentration derivative $\partial a_{\rm X}/\partial c$
rather than the so-called ``Darken-factor'' $\partial \ln a_{\rm
X}/\partial \ln c$ usually written. We also derive equations
describing diffusion of interstitial atoms X in a multicomponent
alloy (Me$_1$Me$_2$...)-X.

In Sec. 4 we  apply the results of Sec. 3 to microscopically treat
the  problem of diffusion of carbon in austenite discussed by a
number of authors {\cite{Agren-82a,Agren-82c,Agren-86,Bhadeshia-81}.
Our treatment is based on the microscopic model of C-C interactions
in austenite suggested by Blanter \cite{Blanter-99} which supposes a
strong ``chemical'' repulsion at short  C-C distances $R$ and a
purely deformational  interaction at longer  $R$. To estimate the
configurational interactions $v(R)$ which determine the carbon
activity $a_{\rm C}$,\, and the ``saddle-point'' interactions
$\Delta(R)$ which determine the above-mentioned correlator \,$b_{\rm
X}$=$b_{\rm C}$,\, we use some plausible assumptions about the
dependences $v(R)$ and $\Delta(R)$ which include adjustable
parameters. The interaction models obtained enable us to describe
both the thermodynamic and the diffusional properties of carbon in
austenite at the same level of accuracy as that achieved in
phenomenological treatments
\cite{Agren-82a,Agren-82c,Agren-86,Bhadeshia-81,Agren-79}. At the
same time, the microscopic approach used enables us to make a number
of qualitative conclusions about the carbon-carbon interactions and
the characteristics of diffusion of carbon in austenite, in
particular, about the presence of a significant C-C attraction at
intermediate $R$ and about a rather low value of  the ``transition
state entropy'' $\Delta S^{SP}_{\rm C}$ given by estimate
(\ref{Delta-S_C}).

\

The authors are much indebted to F. Soisson  for numerous valuable
discussions, as well as to G. F. Syrykh, for drawing our attention
to the important experiments \cite{Sumin-90}.  The work was
supported by the Russian Fund of Basic Research (grant No.
09-02-00563); by the fund for support of leading scientific schools
of Russia  (grant No. NS-7235.2010.2); and by the program of Russian
university scientific potential development (grant  No. 2.1.1/4540).


\begin{thebibliography}{99}

\bibitem{Agren-82a} J. {\AA}gren, J. Phys. Chem. Solids {\bf 43}, 385 (1982).

\bibitem{Agren-82c} J. {\AA}gren, Acta Metall. {\bf 30}, 841 (1982).

\bibitem{Agren-86} J. {\AA}gren, Scripta Metall. {\bf 20}, 1507 (1986).

\bibitem{Bhadeshia-81} H. K. D. H. Bhadeshia, Met. Sci. {\bf 15}, 477 (1981).

\bibitem{Thibaux-07} P. Thibaux, A. Metenier, and C. Xhoffer.  Met. Mater. Trans. A,
{\bf 38}, 1169 (2007).

\bibitem{Onsager-31} L. Onsager,  Phys. Rev. B {\bf 37}, 405 (1931);
{\bf 38}, 2265 (1931).

\bibitem{Martin-90} G. Martin,  Phys. Rev. B  {\bf 41},
2279 (1990).

\bibitem{VBD-95} V. G. Vaks,  S. V. Beiden, and V. Yu.  Dobretsov, Pis. Zh. Eksp. Teor. Fiz.
{\bf 61}, 65  (1995) [JETP Lett. {\bf 61}, 68  (1995)].

\bibitem{Vaks-96} V.G. Vaks, Pis. Zh. Eksp. Teor. Fiz. {\bf 63},
447  (1996) [JETP Lett.  {\bf 63},  471  (1996)].

\bibitem{BV-98} K.D. Belashchenko and V.G. Vaks, J. Phys.: Condensed Matter
{\bf 10}, 1965 (1998).

\bibitem{Nastar-00} M. Nastar, V.Yu. Dobretsov and G. Martin,
Phil. Mag. A  {\bf 80}, 155 (2000).

\bibitem{Vaks-04} V. G. Vaks, Phys. Reports {\bf 391}, 157-242 (2004).

\bibitem{Nastar-11} M. Nastar, Solid State Phenomena
{\bf 172-174}, 321 (2011).

\bibitem{SF-07} F. Soisson and C.-C. Fu,  Phys. Rev. B  {\bf 76},
214102 (2007).

\bibitem{KSSV-11}  K.Yu. Khromov, F. Soisson,  A.Yu. Stroev
and  V.G. Vaks, Zh. Exp. Teor. Fiz. {\bf 139}, 479 (2011) [JETP {\bf
112}, 415 (2011)].

\bibitem{Blanter-99} M. S. Blanter, J. Alloys Comp. {\bf 291}, 167
(1999).

\bibitem{VKh-1}  V. G.  Vaks and K. Yu. Khromov, Zh. Exp. Teor. Fiz.  {\bf 133}, 115 (2008)
[JETP {\bf 106}, 94 (2008)].

\bibitem{VKh-2}  V. G.  Vaks and K. Yu. Khromov, Zh. Exp. Teor. Fiz.  {\bf 133}, 313 (2008)
 [JETP {\bf 106}, 265 (2008)].

\bibitem{Jiang-03} D. E. Jiang and E. A. Carter,  Phys. Rev. B  {\bf 67},
214103 (2003).

\bibitem{Domain-04} C. Domain, C. S. Becquart and J. Foct,  Phys. Rev. B  {
\bf 69}, 144112 (2004).

\bibitem{VZhKh-10} V.G.  Vaks, I.A. Zhuravlev and K.Yu. Khromov,  Zh. Exp. Teor. Fiz.
{\bf 138}, 902 (2010) [JETP 111 (2010) 796].

\bibitem{LL} L. D.  Landau, E. M. Lifshits, {\it Statistical
Physics} (Nauka, Moscow, 1995).

\bibitem {Kittel} C. Kittel, {\it Intruduction to Solid State Physics}
(Wiley, New York, 1953; Nauka, Moscow, 1978), Chap. 6.


\bibitem{VK-93}   V.G. Vaks and V.V. Kamyshenko, Phys. Lett. A {\bf 177}, 269 (1993).

\bibitem{VB-94}   V.G. Vaks and S.V. Beiden,  Zh. Eksp. Teor. Fiz.
{\bf 105}, 1017 (1994)  [JETP {\bf 78}, 546 (1994)].

\bibitem{LBS-02} Y. Le Bouar and F. Soisson,  Phys. Rev. B {\bf 65},
094103 (2002).

\bibitem{Le_Clair-70} A. D. Le Clair,  {\it Correlation Effects in Difusion in
Soilds}, in {\it Physical Chemistry: An Advanced Treatise}, ed. by
H. Eyring (Academic, New York, 1970), Vol. 10, p. 261.

\bibitem{Soisson-08} F. Soisson, private communication.

\bibitem{VKh-1}  V. G.  Vaks and K. Yu. Khromov, Zh. Exp. Teor. Fiz.  {\bf 133}, 115 (2008)
[JETP {\bf 106}, 94 (2008)].

\bibitem{Miller-03} M. K. Miller, B. D. Wirth and G. R. Odette,
Mater. Sci. Eng. {\bf 353}, 133 (2003).

\bibitem{Rana-07} R. Rana, W. Bleck, S.B. Singh, O.N. Mohanti,
Mater. Lett. {\bf 61}, 2919 (2007).

\bibitem{KWZS-08} R.P. Kolli, R.M. Wojes, S. Zaucha and D.N.
Seidman, Int. J. Mat. Res. {\bf 99}, 513 (2008).


\bibitem{VKh-5}  V. G.  Vaks and K. Yu. Khromov, Zh. Exp. Teor. Fiz.  {\bf 136}, 722 (2009)
[JETP {\bf 109}, 619 (2009)].


\bibitem{Satya-85} S. K. Satyia, R. P. Comes and G. Shirane,  Phys. Rev. B  {\bf 32},
3309 (1985).

\bibitem{Neuhaus-97} J. Neuhaus, W. Petry, A. Krimmel, Physica B
{\bf 234-236}, 897 (1997).

\bibitem{Agren-79} J. {\AA}gren, Metall. Trans. A {\bf 10}, 1847 (1979).

\bibitem{Blanter-78} J. {\AA}gren, Metall. Trans. A {\bf 9}, 753 (1978).

\bibitem{Wells-50} C. Wells, W. Batz and R. F. Mehl, Trans. AIME,
J. Metals {\bf 188}, 553 (1950).

\bibitem{Smith-53} R. P. Smith, Acta Metall. {\bf 1}, 578 (1953).

\bibitem{Kurdyumov} G. V. Kurdyumov, L. M. Utevsky, and R. I. Entin,
{\it Transformations in Iron and Steel} (Nauka, Moscow 1977, Chap.
3) [in Russian].


\bibitem{Sumin-90} V. V. Sumin, M. G. Zemlyanov, L. M. Kaputkina,
P. P. Parshin, S. D. Prokoshkin, A. I. Choklo, Fizika Metallov
Metallovedenie,  11, 122 (1990).

\bibitem{Stassis-87} J. Zaretsky and C. Stassis, Phys. Rev. B {\bf 35},
4500 (1987).




\end{thebibliography}
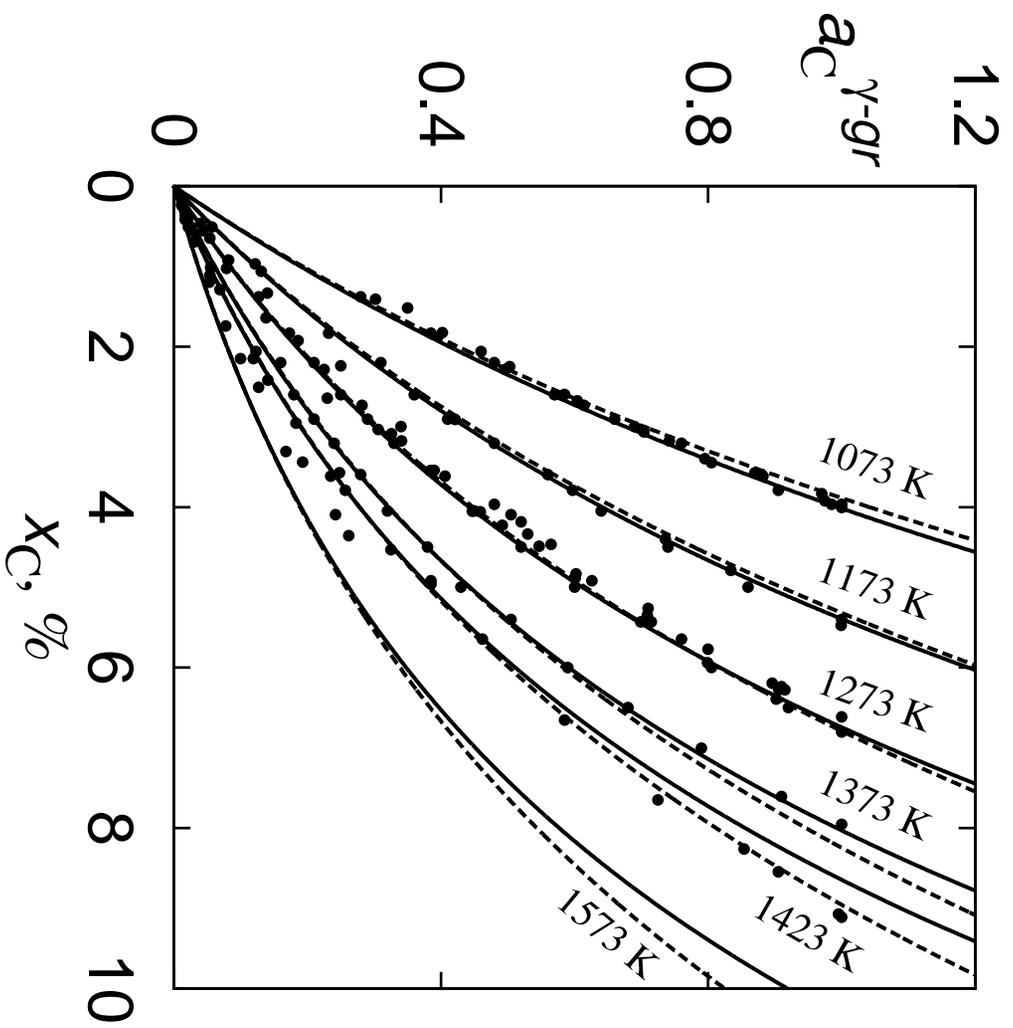
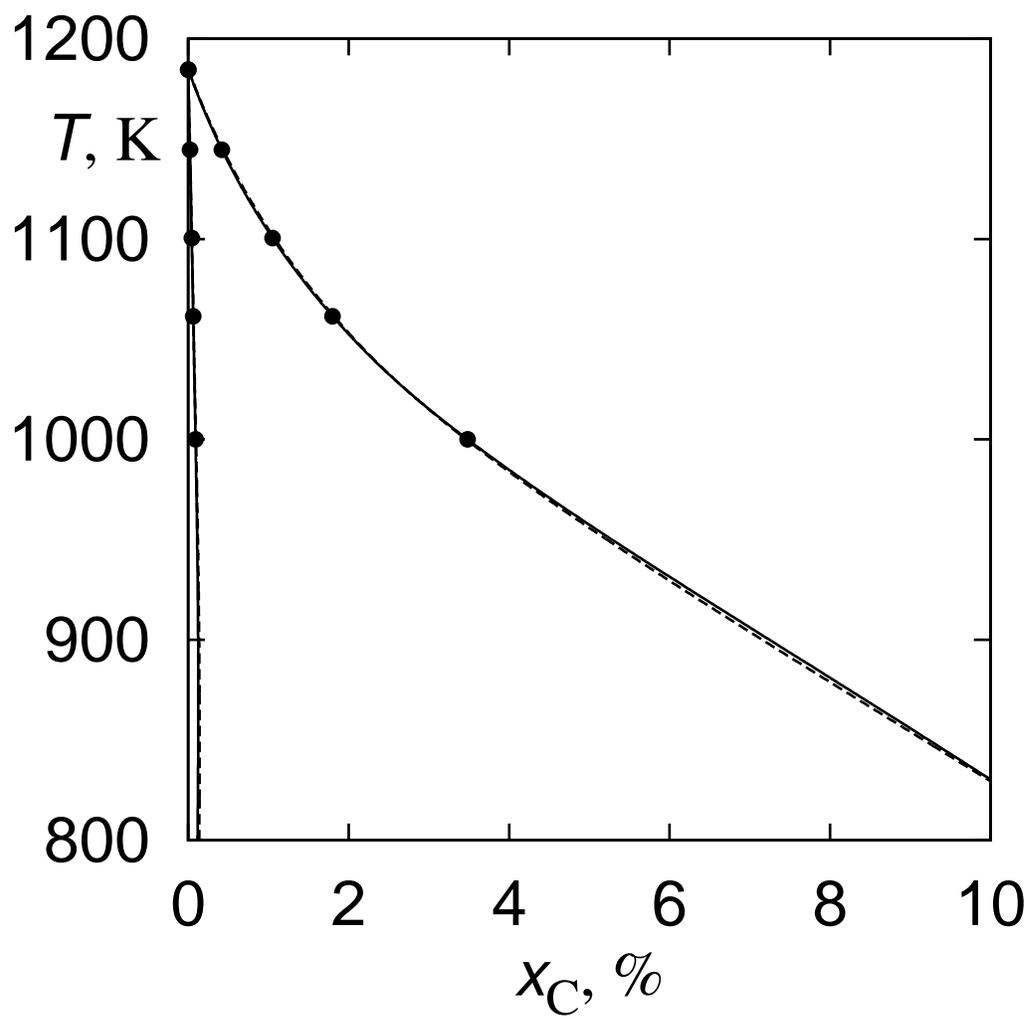
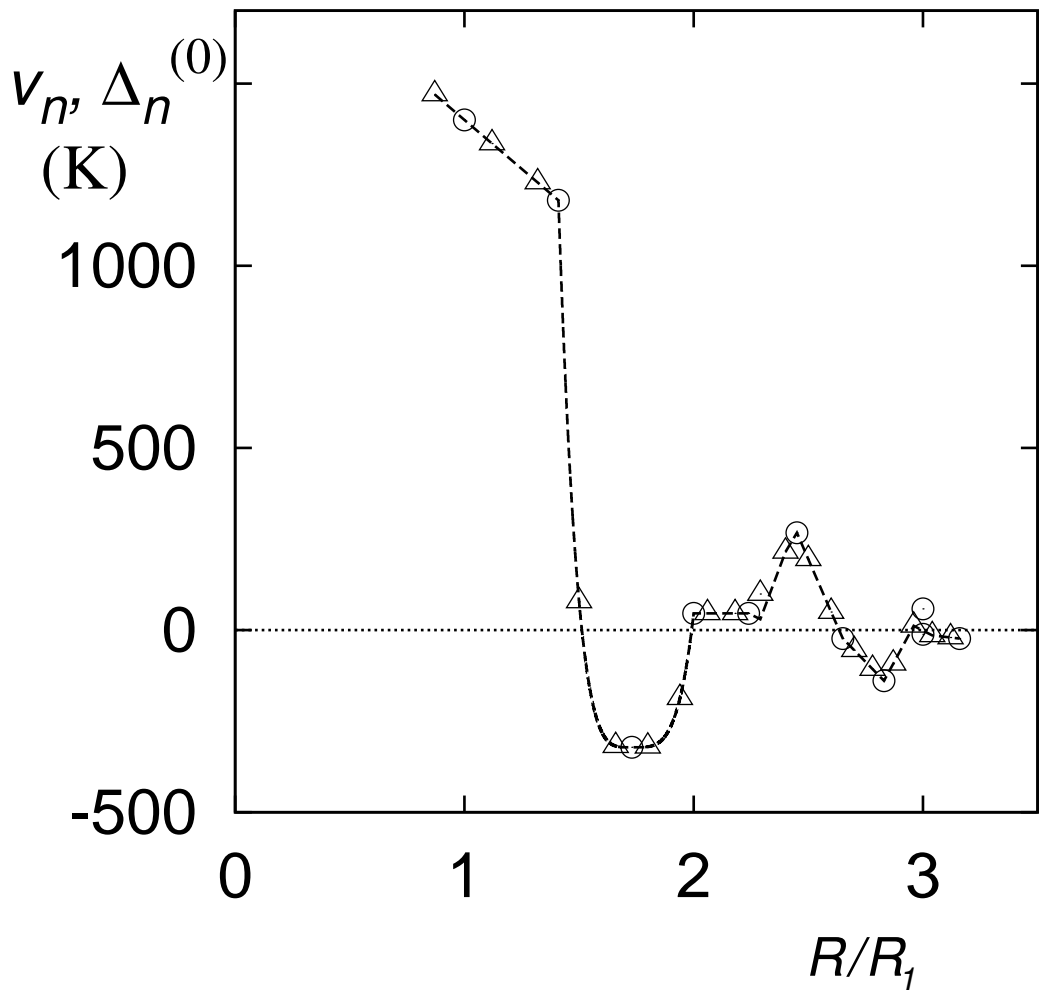
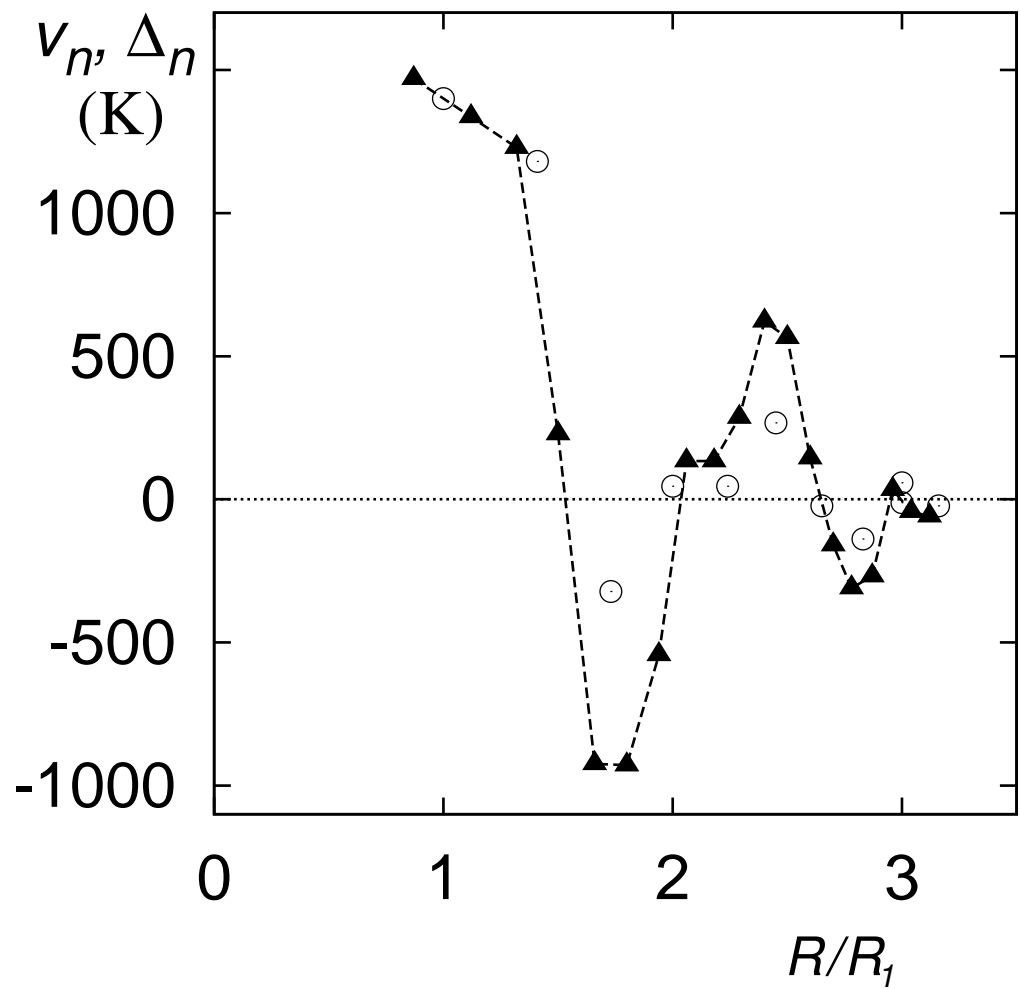
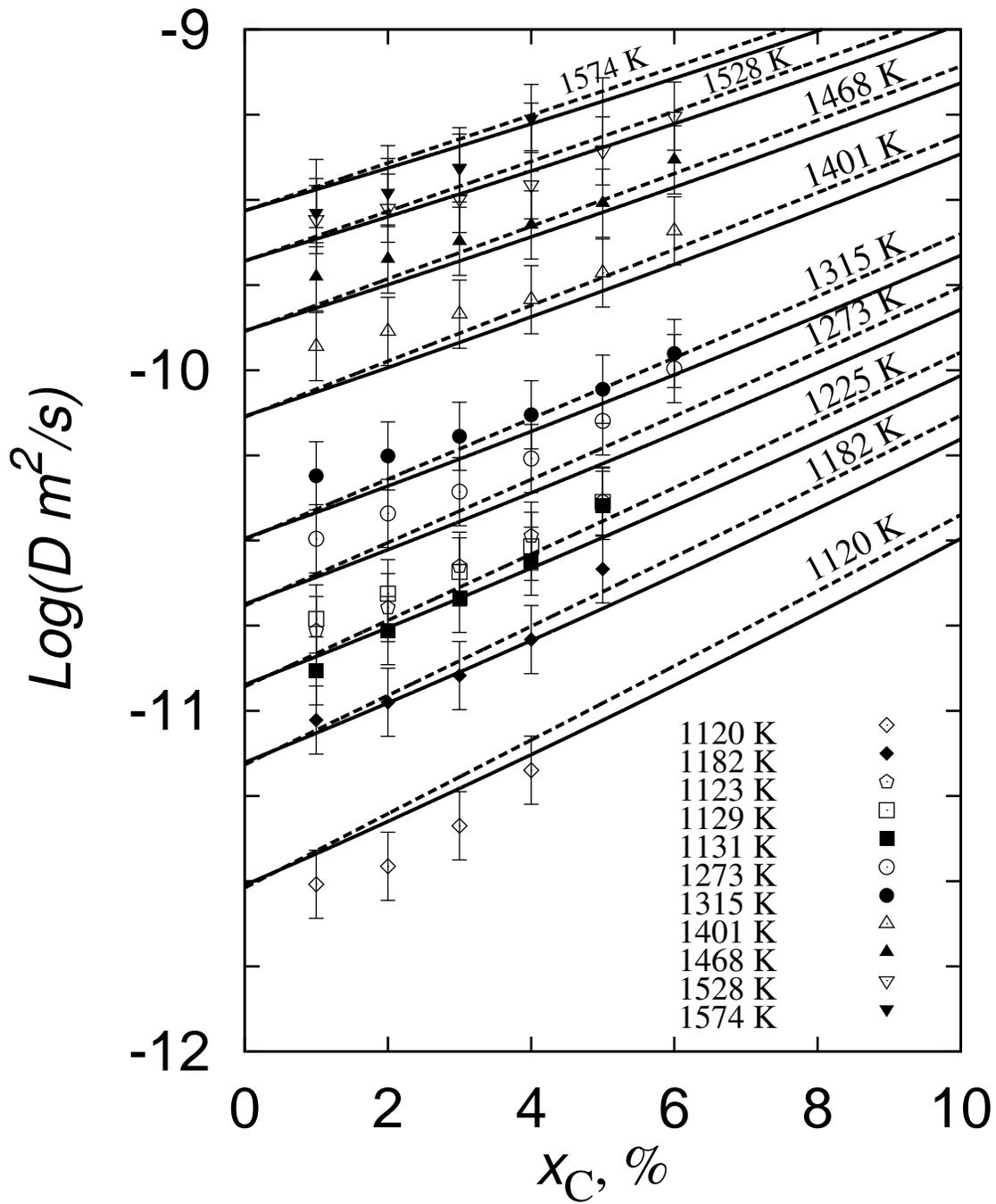
\end{document}